\documentclass{aastex63}

\usepackage{amsmath}

\newcommand{\kms}{km\,s$^{-1}$}
\newcommand{\Halpha}{H$\alpha$}
\newcommand{\Msun}{$M_{\odot}$}

\newcommand{\Vmax}{$V_{\rm max}$}
\newcommand{\Msunpc}{\Msun\,pc$^{-2}$}
\newcommand{\HI}{HI}

\newcommand{\Nsp}{138~}
\newcommand{\Nnir}{18~}

\received{June 00, 0000}
\revised{June 00, 0000}
\accepted{\today}

\shorttitle{Spectra of Superthin Disk Galaxies}
\shortauthors{Bizyaev et al.}

\graphicspath{{./}}

\begin{document}

\correspondingauthor{Dmitry Bizyaev}
\email{dmbiz@apo.nmsu.edu}

\title{Spectral Observations of Superthin Galaxies}

\author{Dmitry Bizyaev}
\affiliation{Apache Point Observatory and New Mexico State University, Sunspot, NM, 88349, USA}
\affiliation{Sternberg Astronomical Institute, Moscow State University, Universitetskiy prosp. 13, 119992, Moscow, Russia}
\affiliation{Special Astrophysical Observatory, Russian Academy of Sciences, 369167 Nizhnij Arkhyz, Russia}

\author{D. I. Makarov}
\affiliation{Special Astrophysical Observatory, Russian Academy of Sciences, 369167 Nizhnij Arkhyz, Russia}

\author{V. P. Reshetnikov}
\affiliation{St.Petersburg State University, 7/9 Universitetskaya nab., St.Petersburg, 199034 Russia}
\affiliation{Special Astrophysical Observatory, Russian Academy of Sciences, 369167 Nizhnij Arkhyz, Russia}

\author{A. V. Mosenkov}
\affiliation{Central Astronomical Observatory, Russian Academy of Sciences, 65/1 
Pulkovskoye chaussee, St.Petersburg, 196140 Russia}

\author{S. J. Kautsch}
\affiliation{Nova Southeastern University, Fort Lauderdale, FL, 33314, USA}

\author{A. V. Antipova}
\affiliation{Special Astrophysical Observatory, Russian Academy of Sciences, 369167 Nizhnij Arkhyz, Russia}



\begin{abstract}

We conduct spectral observations of \Nsp superthin galaxies (STGs) with high 
radial-to-vertical stellar disk scales ratio with the Dual 
Imaging Spectrograph (DIS) on the 3.5m telescope at the Apache Point 
Observatory (APO) to obtain the ionized gas rotation curves 
with R $\sim$ 5000 resolution. 
We also performed near infrared (NIR) H and Ks
photometry for \Nnir galaxies with the NICFPS camera on
the 3.5m telescope.  
The spectra, the NIR photometry and published optical and NIR photometry are used for 
modeling that utilizes the thickness of the stellar disk
and rotation curves simultaneously. The projection and dust extinction effects are taken 
into account.  
We evaluate eight models that differ by their 
free parameters and constraints. 
As a result, we estimated masses and scale lengths of the galactic dark halos. 
We find systematic differences between the properties of our red and blue STGs.
The blue STGs have a large fraction
of dynamically under-evolved galaxies whose vertical velocity dispersion
is low in both gas and stellar disks.
The dark halo-to-disk scale ratio is shorter in the red STGs than in the blue ones,
but in a majority of all STGs this ratio is under 2.
The optical color $(r-i)$ of the superthin galaxies correlates with 
their rotation curve maximum, vertical velocity dispersion in 
stellar disks, and mass of the dark halo.   
We conclude that there is a threshold central
surface density of 50 $M_{\odot}$\,pc$^{-2}$ below which we do not observe
very thin, rotationally supported galactic disks.

\end{abstract}


\section{Introduction}

The thinnest disk galaxies have been studied since a long time because of their unusual appearance, e.g., \citet{VV67,goadroberts79,FGC,kautsch06}. 
These superthin galaxies (STGs) are traditionally defined to have major-to-minor axial ratios larger than nine ($a/b>9$) when seen edge-on \citep{goadroberts79,goadroberts81}. 
These galaxies are a subgroup of flat disk galaxies and are rare \citep{rfgc,makarov01,mitronova05,kautsch06,EGIS}. 
\citet{B17} (B17 hereafter) found their fraction is almost 4 per cent among the late-type morphological galaxy class Sd, 
and 5 per cent among the Sd class in the \citet{kautsch06} catalogue. 
Hence, the STGs represent the `faint end' of `flat' and very thin disk galaxies analyzed and reviewed by \citet{FGC,rfgc,kautsch09}.

The main mystery of STGs is their appearance. 
It is unclear what processes create these thin stellar disks and subsequently what evolution preserves the disks from morphological distortion. Under normal circumstances, galactic stellar disks are under the influence of efficient disk heating processes. 
These external and internal processes increase the disk thickness and produce the "regular" morphology of the majority of today’s disk galaxies \citep{kormendy83,naab03,sotnikova03,kormendy05,donghia06,bournaud07,yoachim08,bullock08,kazantzidis08,kazantzidis09,purcell09,kh10}. 
In this article, we focus on spectroscopic observations of STGs and how those may help explain the resistance of STGs against morphological transformations and disk thickening.

Only a few nearby superthin galaxies have been studied in detail so far. Multi-wavelength data obtained for a `prototype' STG UGC~7321 \citep{matthews99,matthews00,matthews01,uson03} allowed authors to conclude that this galaxy contains a large fraction of dark matter and a low surface brightness (LSB) stellar disk. 
\citet{kautsch09,B17} came to conclusions that the STGs are LSB galaxies. 

It appears that the majority of STGs and simple disk galaxies is located in quiet environments, which may prevent them from external perturbations \citep{kautsch09b}. 
B17 noticed that small STGs avoid cosmological filaments, which indicates that these blue, 
under-evolved galaxies avoid the high density environment. The latter is in agreement with similar conclusions made for LSB galaxies by \citet{rosenbaum04,rosenbaum09}. 
\citet{karachentsev16} discovered that flat galaxies with very thin disks have fewer satellite companions than regular spiral galaxies. 
Moreover, spectroscopic studies of small samples of superthin galaxies \citep{goadroberts81,vdKruit01,matthews08,obrien10,B17,banerjee17,kurapati18}
indicate that the galaxies grew up and evolved in rather unperturbed cosmological environments (B17) and hence, allow us to study the galactic disk formation disentangled from the consequent dynamical evolution.

All current considerations about STGs point to the conclusion that dark matter (DM) plays the dominant role in shaping these objects and keeps them stable. \citet{banerjee17} noticed that special structural features of dark matter halos in STGs like their compactness, may be responsible for the formation of their thin, LSB disks. 

Recent studies found evidence that massive dark halos can press the galactic stellar disks of STGs into their thin shape with low stellar volume density. 
\citet{zasov91,zasov02,kregel05b,sotnikova06,kh10}
found a correlation between the relative disk thickness and the relative mass of a spheroidal DM halo. 
Other numerical simulations demonstrate that LSB galaxies prefer to form in dark matter halos with long spatial scales \citep{maccio07}. 
Moreover, \citet{dipaolo19} studied 72 LSB galaxies and found that the extension of stellar disks (i.e., their disk scale length) increases with larger dark halo scale lengths. 
Also B17 suggest a rather long dark halo scale lengths for the STGs, in agreement with \citet{maccio07}. 
At the same time, \citet{banerjee13,banerjee17,kurapati18} conclude that dark halos of a few studied STGs have short scale length if expressed in the stellar disk scales. 

\citet{banerjee17,kurapati18} 
concluded that 
the halo compactness seems to have an important role in shaping the stellar disks in STGs,
although the sample they worked with was very limited.
Simulations by \citet{dipaolo19} showed that the rotational velocity of the stellar disks 
is proportional to the halo-to-disk scale ratio. 
This means that a stellar disk rotates slow when its DM halo is small and compact, and the stellar disk is thin and extended. 
\citet{kormendy16} noticed that the dark matter halos in late-type spiral and dwarf spheroidal galaxies have smaller core radii and higher central densities, 
which means their dark halos are compact. 
In contrast, observations of nearby LSB galaxies indicate that the dark matter halo profiles are cored rather than compact ("cusped") \citep{kuziodenaray11, salucci12}, suggesting long dark halo scales.  
\citet{kormendy04,kormendy16,dipaolo19} found that the central surface density of the dark matter halos is constant in a very wide range of objects 
with various absolute magnitudes ($M_B$ from $-5$ to $-22$~mag). 
In this case, the only feature of the dark halos that affects the evolution of the baryonic matter is the scale length of the dark halo (setting aside the asymmetry and interaction between different halos).



The further investigation of possible connections between the properties of the dark halos and the stellar disks should give 
a valuable feedback to disentangle those discrepancies and feed modern cosmological simulations. 
Particularly, many STGs are found in unperturbed conditions that did not change significantly over the lifetime 
and thus provide a well-defined test pad for those simulations. 
A large number of STGs is required to make reliable, statistically valuable conclusions. 

Public surveys provide us with a way to select the STGs, but the spectroscopic data are insufficient 
to study a large STG sample. 
Here we rely on a sample of verified edge-on galaxies by~\citet{EGIS} 
(`Edge-on disk Galaxies In SDSS', 
hereafter EGIS\footnote{\url{http://users.apo.nmsu.edu/~dmbiz/EGIS/} \newline \url{https://www.sao.ru/edgeon/catalogs.php?cat=EGIS}})
A part of these galaxies was observed in the near-infrared (NIR) by \citet{nir}. 
In this paper we report results from our long-term spectroscopic campaign, as well as an additional NIR photometry.

The cosmological framework adopted throughout this paper 
is $H_0$ = 72 km s$^{-1}$ Mpc$^{-1}$ , $\Omega_m$ = 0.3, and $\Omega_\Lambda$ = 0.7.

\section{The Sample of superthin edge-on galaxies}

Our sample of superthin galaxies was selected using the same criteria described in \citet{B17}.
We used the EGIS \citep{EGIS} catalogue to select galaxies with $h/z_0 > 9$ in the $r$-band.
We also added several galaxies with visually thin disks with the bluest colors among the EGIS sample, 
even if their thickness did not meet the $h/z_0$ criterion. 

We targeted some of selected galaxies for NIR photometric observations (see below) and attempted to observe
more objects to enhance the NIR sample in coordination with the sample by \citet{nir}.

\subsection{Spectral Observations of Superthin Galaxies}

We continued observations in the same manner as described by \citet{B17}.
We were able to collect spectra of \Nsp objects with the Dual Imaging Spectrograph
(DIS) on the 3.5m telescope at the Apache Point Observatory (APO) between 
December 2014 and August 2019. All observations were conducted in
the high-resolution mode (B1200/R1200 grating),
which provides the spectra resolution of 5000.
In addition to the H$\alpha$ emission, we can see other
major emission lines 
(H$\beta$, [OIII]4959,5007\AA, [NII]6548,6583\AA, \& [SII]6717,6731\AA)
in about three quarter of all galaxies.
The deterioration of blue DIS spectra after 2017 because of condensation
in the blue DIS dewar made the blue spectra less useful for kinematic studies.

Similar to the observations described by B17, each galaxy was 
observed with one to three exposures, from 15 to 60 minutes long. 
A Helium-Neon-Argone wavelength calibration lamp was observed immediately
after each galaxy at the same position in the sky. Spectrophotometric 
standard stars were observed every night.
Each observing night was accompanied with a set of biases and dome flats.
The data reduction was performed with IRAF standard tools, including
bias subtraction, flat fielding, wavelength calibration, sky line
and sky background subtraction, cosmic ray removal, and flux calibration.
Same as in \citet{B17}, the typical accuracy of the radial velocity is 12~\kms.

Table~\ref{tab:tab1} shows the object name (in the EGIS catalogue), date of observations, 
total exposure time, and our heliocentric radial velocity in \kms.

\begin{table}
\centering
\caption{Spectral Observations of Superthin Galaxies.
Name, date of observations, total exposure time, and heliocentric radial velocity (\kms).
Note that the names of the galaxies from the EGIS catalogue contain RA and Dec coordinates of their center in decimal degrees.  
\label{tab:tab1}}
\begin{tabular}{llll}
\hline
Object & Date & Exposure & RV \\
            &         &      min             &    \kms{}           \\
\hline
EON\_105.825\_13.464  & 31 Jan 2016  & 40  &   20373  \\
EON\_11.160\_-11.189  & 29 Aug 2019  & 25  &    8106  \\
EON\_115.887\_31.535  & 01 Feb 2019  & 15  &    3750  \\
EON\_116.146\_18.328  & 31 Jan 2016  & 35  &   14604  \\
EON\_118.918\_28.743  & 01 Feb 2019  & 12  &    6434  \\
\multicolumn{4}{c}{... Table is published in its entirety}\\
\multicolumn{4}{c}{in the electronic edition. ...}\\
\hline
\end{tabular}
\setcounter{table}{1}\\
\end{table}

\subsection{Near Infrared Photometry of Superthin Galaxies}
\label{Snir}

The paper \citet{nir} describes near infrared photometry of 49 galaxies from our list. 
We conducted additional observations of \Nnir galaxies with the NICFPS camera 
on the 3.5m telescope at the Apache Point Observatory (APO). 
Each galaxy was observed in a sequence of 20 to 60 short dithered exposures
that prevented the center of galaxies saturate, and gave sufficient signal
to combine the images for estimating the sky background. 
We ensured that the overall exposure time is sufficient for providing
with images of the galaxies visually extended similar to corresponding
SDSS images. 
Flat fielding and image 
co-adding was done using the IRAF standard packages.
Similar to \citet{nir}, we applied photometric calibration of the images
via non-saturated stars in the fields identified also in the 2MASS survey. 
Typically we used seven stars in each field and applied the 2MASS-MKO transformation 
from \citet{legget06}. 
The NIR observations are summarized in Table~2.

\begin{table}
\centering
\caption{Objects, Date of observations, Filter, Exposure time
and seeing FWHM in arccsec.
\label{tabnir}
}
\begin{tabular}{lllll} 
\hline 
Object &  Date & F & Exp. &  FWHM \\
\hline
\hline
EON\_119.500\_17.903  &  02 Dec 2015 & H & 40x25 sec & 1.61 \\
EON\_123.495\_45.742  &  02 Dec 2015 & H & 40x25 sec & 0.93 \\
EON\_128.304\_51.316  &  02 Dec 2015 & H & 40x15 sec & 1.17 \\
EON\_129.751\_27.821  &  02 Dec 2015 & H & 40x25 sec & 1.09 \\
EON\_132.574\_3.497   &  02 Dec 2015 & H & 40x30 sec & 1.20 \\
\multicolumn{5}{c}{... Table is published in its entirety}\\
\multicolumn{5}{c}{in the electronic edition. ...}\\
\hline
\end{tabular}
\label{tab:tab2}
\setcounter{table}{2}\\
\end{table}

\section{Properties of Our Galaxies}

\subsection{Observing Radial Velocities and Rotation Curves}

The rotation curves were obtained from the final, wavelength calibrated and sky subtracted spectra. 
The emission line profiles in the spectral direction were fitted by a gaussian. The gaussian 
center position gave us the radial velocity estimate. The uncertainty of the gaussian 
centroid position combined with the uncertainty of the wavelength calibration in quadratures
were used as the uncertainty estimate for individual data points on the rotation curves. 

The obtained rotation curve was folded in the spatial direction. 
The resulting curve was fitted with a third-order polynomial after which 
2 to 3 sigma outliers were removed. 
The spatial position of the center was slightly varied and fitted
to make both sides of the curve show the lowest scatter of the radial velocities.
The procedure was repeated iteratively until finding the best-fit rotation curve.
We also considered to utilize different approximation of the rotation curve shape using equations from \citet{courteau97}, and came to the same resulting observing points on the rotation curves after
removing outliers using the procedure described above. The rotation curve maximum 
$V_{max}$ is determined as the highest data point value on the observing rotation curve.

\begin{figure}
\epsscale{1.00}
\plotone{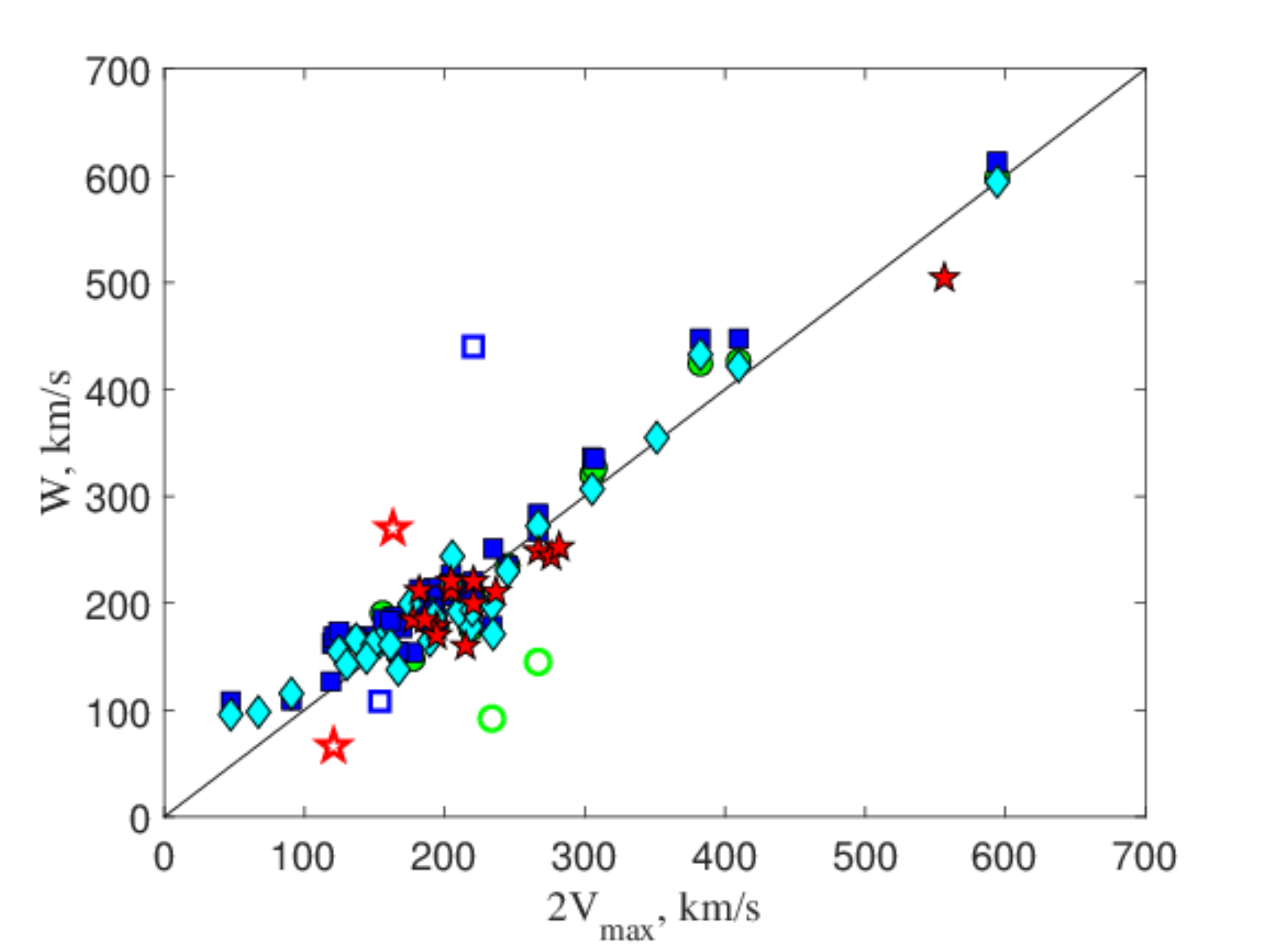}
    \caption{Comparison of our \Vmax{} with published widths of 
    radiolines by their 50 per cent level $W_{50}$. 
    The green circles designate the objects from \citet{springob05},
    the blue squares correspond to \citet{EDD},
    the cyan diamonds show sample from \citet{ALFALFA},
    and the red stars represent direct rotational velocity maxima from published 
    rotation curves (see text). 
    The data points rejected from the analysis are designated by open symbols.
    The solid black line corresponds to the relation $W_{50} = 2 V_{\rm max}$. }
    \label{fig:VmaxW}
\end{figure}

HyperLeda database\footnote{\url{http://leda.univ-lyon1.fr}} \citep{HyperLeda} 
contains radial velocity data for many galaxies from our spectral sample.
There are 98 optical redshift measurements for 71 our galaxies. 
52 of them were reported by SDSS survey \citep{sdssRV}.
The mean radial velocity difference (after rejection of some highly deviating points) 
is $-1.4\pm2.9$~\kms{} with a standard deviation of 27~\kms, which indicates an excellent agreement.
In addition, \HI-observations are available for 52 galaxies from our sample.
The data are taken from 
the \HI-catalogue of the Extragalactic Distance Database EDD by \citet{EDD} (40 galaxies),
from the ALFALFA survey by \citet{ALFALFA} (31 objects) and from \citet{springob05} (24 galaxies).
\HI-observations of edge-on spirals by \citet{giovanelli97} were performed for 14 galaxies
and \HI{}-survey of highly flattened, edge-on, pure disk galaxies 
by \citet{matthews00b} gave us data for comparison for 13 objects.
After rejecting a few outliers the mean difference between \HI-data and the systematic 
velocity from the rotation curve is $7.9\pm2.3$~\kms{} with a standard deviation of 29~\kms.
Having in mind methodological differences of systemic velocity determination in 
\HI{}  and long-slit observations, we conclude that our data are in a good agreement with 
radio observations.

Thirty-nine galaxies from the spectral list do not have radial velocities in the NED database\footnote{\url{https://ned.ipac.caltech.edu}}. 
Sixteen more those galaxies have NED radial velocities different by more than 100~\kms{} from those in Table~\ref{tab:tab1}.
Thus, we report 40 per cent of new, correct radial velocities for our sample of superthin galaxies.
The mean radial velocity difference between the NED and Table~\ref{tab:tab1} for the other 84 galaxies is $-1.4$~\kms{} 
with the sigma of 39~\kms. 
The median difference and the MAD scatter are even lower: $-0.4$ and 31~\kms, respectively.  

Besides the redshifts, the HyperLeda database provides information about kinematics of galaxies 
as \HI-line-width measurements and the rotation curves maximum.
Fig.~\ref{fig:VmaxW} shows comparison of our \Vmax{} measurements with published 
data on the \HI-line-width by its 50 per cent level $W_{50}$:
24 galaxies from \citet{springob05} are designated with green circles,
42 blue squares correspond to the \HI{} data by \citet{EDD},
31 cyan diamonds show the data from ALFALFA survey \citep{ALFALFA},
and 17 red stars for 14 different galaxies represent direct rotational velocity maximum 
measurements from rotation curves. 
The rejected data points are designated by open symbols.
The black line marks the $W_{50}=2V_{\rm max}$ correspondence.
Table~\ref{tab:VmaxW} summarises results of a linear regression between our measurements of \Vmax{}
and literature data on $W_{50}$ in the form of $V_{\rm max}=a+b (W_{50}/2)$.
The zero-point $a$ is essentially zero, while the slope $b$ is 
equal to one, statistically.

\begin{table}
\centering
\caption{Comparison source, number of common galaxies N, linear regression parameters
$a$ and $b$ for $V_{\rm max}=a+b (W_{50}/2)$, and the regression residuals $\sigma_V$.
Designations of the sources follows: 
S05 - \citet{springob05}, EDD - \citet{EDD} and ALFALFA - \citet{ALFALFA}.
The RC refers to the individual data sources mentioned in the paper, see text.
\label{tab:VmaxW}}
\begin{tabular}{lrcll}
\hline
Source & 
N      & 
\multicolumn{1}{c}{a} & 
\multicolumn{1}{c}{b} & 
\multicolumn{1}{c}{$\sigma_V$}  \\
&    
& 
\multicolumn{1}{c}{\kms}      & 
\multicolumn{1}{c}{\kms}      & 
\multicolumn{1}{c}{\kms} \\
\hline
S05 & 22 &  $2.6\pm5.7$ & $0.95\pm0.04$ & 11.1 \\
EDD        & 43 &  $0.8\pm4.7$ & $0.94\pm0.04$ & 11.9 \\
ALFALFA    & 31 &  $0.5\pm5.3$ & $0.98\pm0.04$ & 13.3 \\
RC                 & 15 & $-5.1\pm9.0$ & $1.11\pm0.08$ & 11.5 \\
\hline
\end{tabular}
\end{table}

Data on the rotation curves maximum for a few objects show contradictions between different literature sources 
and our observations. 
We comment on them with possible explanation below. 

EON\_019.768\_-00.139 (PGC~4734, RFGC~296, UGC~847). 
The spectrum of this object by \citet{springob05} 
has low signal-to-noise ratio that led to significant underestimation of its $W_{50}=145$~\kms.

EON\_225.915\_42.126 (PGC~53744, RFGC~2908, UGC~9681). 
This galaxy forms a close pair with PGC~53758 = UGC~9684. 
The angular separation is 2.1 arcmin only.
As a result, all \HI-data are contaminated by the companion.

EON\_032.766\_06.667 (PGC~8353, RFGC~469, UGC~1677).
Most of the radio data for this galaxy are wrongly associated with \HI-emission from a nearby 
galaxy UGC~1670 with $V_h\sim1600$~\kms.
However, ALFALFA survey \citep{ALFALFA} reports reliable \HI-velocity and line-width estimation for our object.

\subsection{Structural Parameters}

In this paper we make use of the structural parameters of galactic disks estimated both in the optical
and NIR bands using SDSS archival photometry processed by \citet{B17} and NIR photometry 
from \citet{nir} and this paper. 
The optical and NIR disk scales for the galaxies were estimated with pipelines described by 
\citet{EGIS} and \citet{nir}. 
The disk scales were estimated taking the spatial resolution of 
optical and NIR photometric observations into account. We ensure that the 
limited spatial resolution of the imaging does not affect the estimated 
vertical scale heights in out galaxies systematically, according to 
the conclusions made by \citet{EGIS}.
In addition to the scales reported in the published papers, we 
estimate the vertical and radial disk scales from the NIR photometry of \Nnir galaxies
reported in this paper. The structural parameters are shown in Table~\ref{tab:tab3}.

Despite the performed NIR observations for some of galaxies from the spectroscopic sample,
most of them do not have the disk scales in the NIR. Fortunately, \citet{nir} showed that 
the optical and NIR scales are connected, and we can estimate the NIR scales using 
SDSS data from  \citet{EGIS} using the following relations:
\begin{equation}
\begin{split}
h(H)  \,=   0.481\,(\pm 0.099)\, h(r),  \\
z_0(H)  = 0.855\,(\pm 0.138)\, z_0(r)
\end{split}
\label{hznir}
\end{equation}

As it was noticed by \citet{nir}, the disk scale height is almost the same in the NIR and optical
measurements, while the near-infrared disk scale length is significantly shorter. 
The latter can not be explained by the dust attenuation effects alone, and reveals a real difference in distribution 
of young and old stellar populations though the galactic disks. 

The central surface brightness is affected by the dust extinction and we were not surprised to see a large
scatter for the relation between the optical $\mu_r$ and NIR $\mu_{H,K}$ central surface brightness. 
Incorporating the optical color $(r-i)_0$ helps reduce the uncertainties of such a cross-calibration. 
We perform a two-dimensional fitting to the data from  \citet{nir} and find the following relation:\\
\begin{equation}
\mu_H \,=\, 12.227  - 7.043\, (r-i)_0  +  0.415\, \mu_r ~~,
\label{munir}
\end{equation}
where the $(r-i)_0$ is the SDSS color of galaxies corrected for the Milky Way extinction from \citet{EGIS}.
Although the relation provides a rather coarse cross-calibration with the uncertainty of 0.7 mag, it is helpful
for including the galaxies without NIR observations into the further analysis,
given we do not use photometric data to convert the surface 
brightness to surface density directly.

The rotation curve maximum, optical and NIR structural parameters are shown in Table~\ref{tab:tab3}. 
The `NIR flag' Fnir designates the source of the NIR 
protometry for the galaxy, and can be `h' or `k' when the APO 3.5m data are used,
`H' or `K' when the data are taken from \citet{nir}, or `F' if equation (\ref{hznir})
was applied. The distances in Table~\ref{tab:tab3} correspond to our radial velocities
corrected for the CMB motion according to the NED\footnote{https://ned.ipac.caltech.edu}.

\begin{table*}
\centering
\caption{Parameters of Superthin Galaxies. The columns show
the name of galaxies, distance in Mpc that corresponds to the radial velocity corrected 
for the CMB motion, the radial scale $h$ and its uncertainty from the EGIS catalogue, in kpc, 
the vertical scale $z_0$ and its uncertainty from the EGIS catalogue, in kpc, the same 
scales and their uncertainties from NIR opbservations (see text), the $(r-i)_0$ color, 
the maximum of the rotation curve velocity, and the "Fnir" flag (see text).}
\begin{tabular}{lllllllllllll}
\hline
Name   &           D,Mpc&  hopt&  dhopt& z0opt&dz0opt& hnir&  dhnir& z0nir&dz0nir& (r-i)&\Vmax&Fnir \\
\hline
EON\_105.825\_13.464 & 285.14 & 7.881 & 1.297 & 0.789 & 0.220 & 3.791 & 0.623 & 0.675 & 0.188 & 0.379 &142 &F\\
EON\_11.160\_-11.189 & 108.10 & 4.368 & 0.069 & 0.889 & 0.266 & 2.101 & 0.034 & 0.760 & 0.227 & 0.232 &131 &F\\
EON\_115.887\_31.535 &  54.43 & 2.892 & 0.018 & 0.520 & 0.076 & 1.391 & 0.009 & 0.445 & 0.065 & 0.149 & 86 &F\\
EON\_116.146\_18.328 & 205.64 & 4.833 & 0.127 & 0.804 & 0.081 & 2.325 & 0.061 & 0.687 & 0.069 & 0.405 &154 &F\\
EON\_118.918\_28.743 &  92.00 & 2.643 & 0.222 & 0.852 & 0.143 & 1.271 & 0.107 & 0.729 & 0.122 & 0.156 & 62 &F\\
\multicolumn{13}{c}{... Table is published in its entirety in the electronic edition. ...}\\
\hline
\end{tabular}
\label{tab:tab3}
\setcounter{table}{3}
\end{table*}

\begin{figure}
\plotone{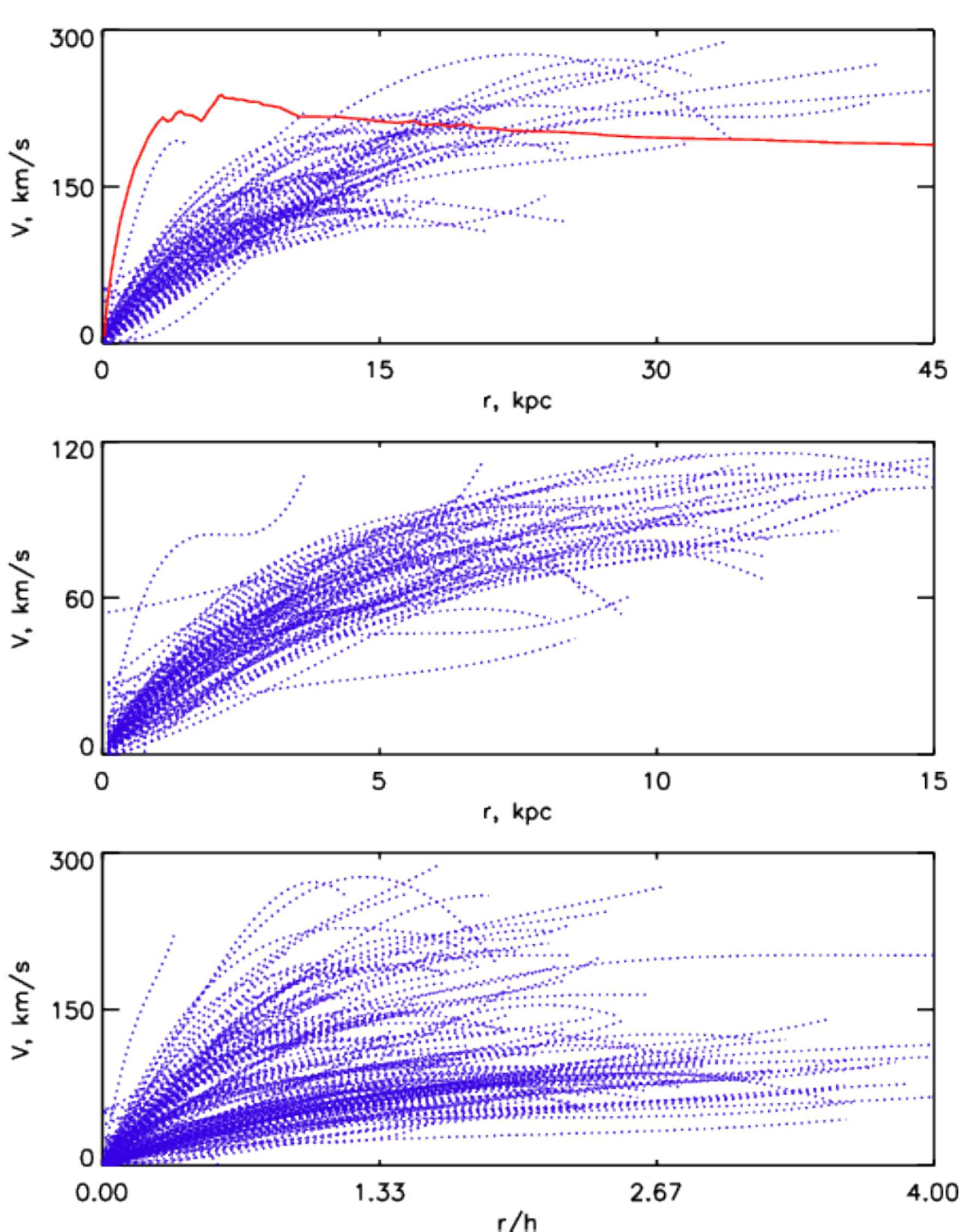}
\caption{Rotation curves of our \Nsp superthin galaxies are shown with the blue dotted curves for objects with $V_{max} > $ 120 \kms ~ (upper panel) and $V_{max} \le $ 120 \kms ~ (middle panel).
The solid red curve demonstrates the Milky Way's rotation curve.
The lower panel repeats the plots above for the same 
rotation curves with radius normalized by the $r$-band disk scale 
length. 
\label{fig1}}
\end{figure}

Fig.~\ref{fig1} demonstrates the rotation curves of all galaxies from Table~\ref{tab:tab1}. The red solid curve shows
the Milky Way galaxy \citep{sofue99}.
The lower panel demonstrates our rotation curves with 
the radius normalized by the disk scale length in the $r$ band.
Fig.~\ref{fig1} reveals that the majority of our superthin galaxies 
have rotation curves with shallow central parts, 
which is an indication of the dominance of DM in STGs.
The rotation curve maximum, if detected, is reached at several
times farther from the center than in the regular galaxies, which is a typical feature
of low surface brightness galaxies.  

\subsection{Properties of Rotation Curves in Superthin Galaxies}

\begin{figure}
\epsscale{1.00}
\plotone{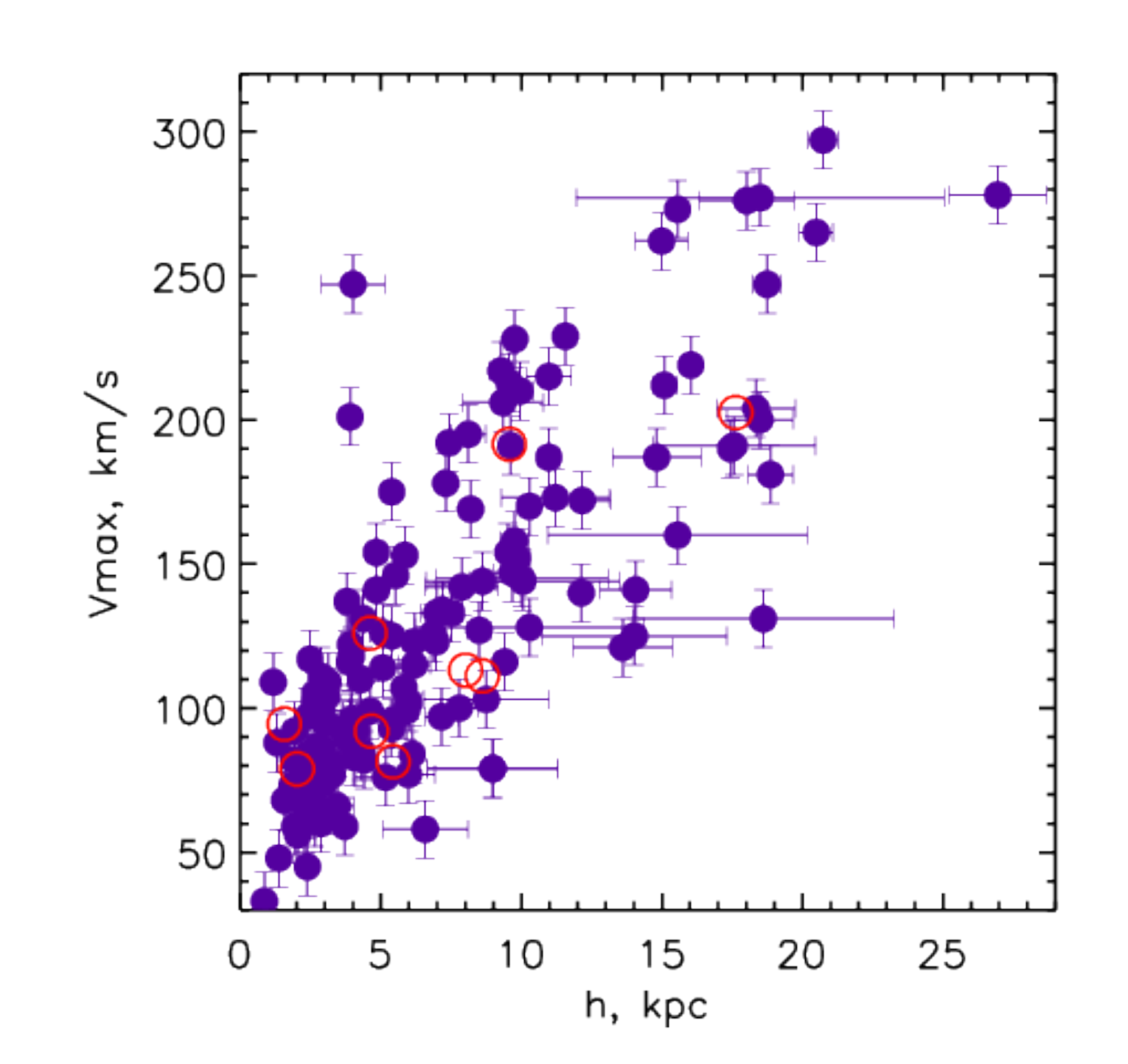}
\caption{
The rotation curve maximum \Vmax{} versus the $r$-band scale length for our (blue bullets) and published (red circles) data. 
\label{fig2}}
\end{figure}

Fig.~\ref{fig2} shows the relation between the disk scale length (estimated in the r-bang here) and the maximum rotation
velocity \Vmax. We confirm the trend reported by B17. 
The robust linear fitting to Fig.~\ref{fig2} gives $h\, \propto \, V_{\rm max}^{1.6}$. 
This result is in a qualitative agreement with similar relations 
found by \citet{courteau07,hall12} made for large samples
of regular disk galaxies. 

\begin{figure}
\epsscale{1.20}
\plotone{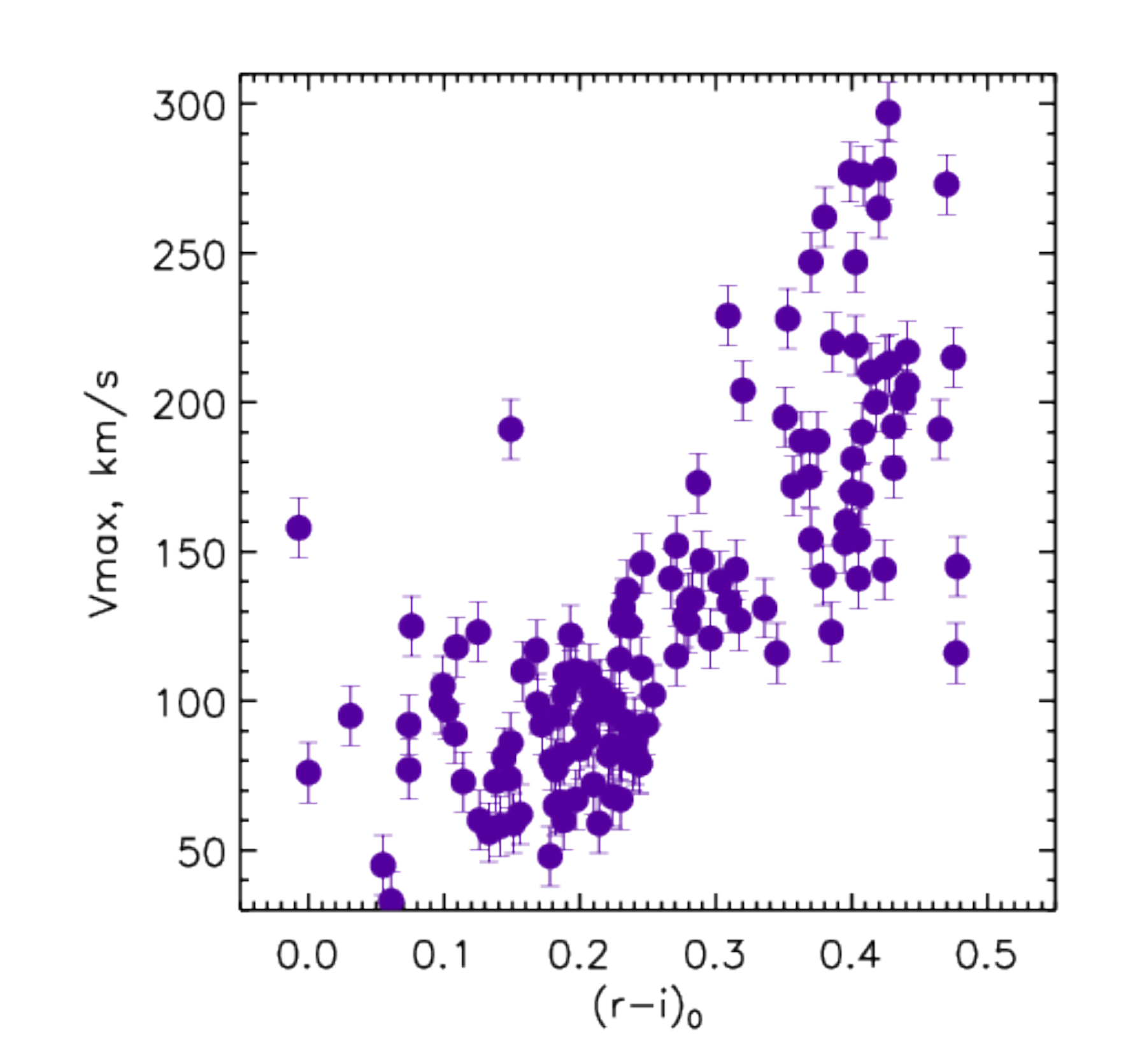}
\caption{The rotation curve maximum \Vmax{} versus the galaxy color 
$(r-i)_0$ corrected for the galactic foreground extinction. 
\label{fig3}}
\end{figure}

According to B17, the STG is a non-uniform group of galaxies that demonstrates different properties between the red and blue objects:
the red galaxies are large and dynamically evolved, while the blue galaxies are small and dynamically cold. Fig.~\ref{fig3} shows
that the maximum rotation velocity \Vmax{} depends on galactic color, which is in agreement with B17. The blue galaxies with $(r-i)_0 \, \le $ 0.24 mag
are mostly low massive, with low \Vmax{} independent of color. The red galaxies with 
$(r-i)_0 \, > $ 0.24 mag  are large, and we can see a 
positive correlation between their \Vmax{} and color. 
Note that the $(r-i)_0 = $ 0.24 mag threshold splits our sample of galaxies in equal halves. 

\begin{figure}
\epsscale{0.6}
\plotone{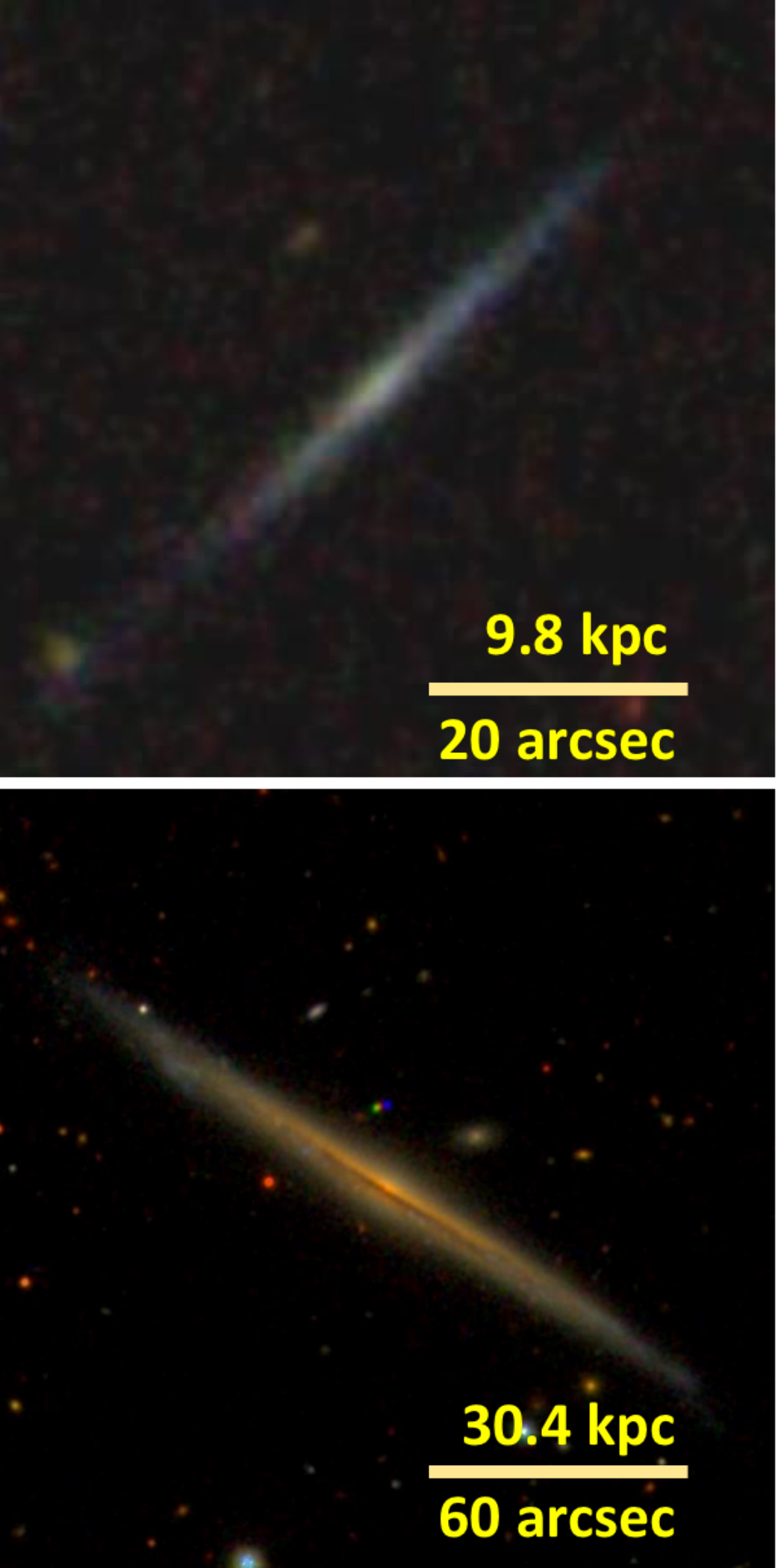}
\caption{Examples of blue (EON\_32.766\_6.667, top) and red (EON\_149.150\_20.646, bottom) superthin galaxies from our sample. The horizontal bar in each 
panel demonstrates the scale in arcseconds and kpc at the distance of the galaxies. 
\label{fig3a}}
\end{figure}

We show examples of blue and red galaxies in Fig.~\ref{fig3a}. The upper panel demonstrates a 
slowly rotating blue STG EON\_32.766\_6.667, 
while the lower panel shows a large, red, massive STG
EON\_149.150\_20.646.

Significant fraction of additional radial velocity measurements allows us to compare the physical
scale length of regular and superthin galaxies, similar to what has been done by B17. We 
confirm the conclusion by B17 that the STGs have bluer colors for the same scale length as regular galaxies. 

\subsection{Stability of Disks and their Thickness}
\label{vrothz}

As it was noticed by B17, a simple galaxy model with an exponential stellar disk and a spherical gravitational potential of halo
enables us to estimate the relative disk thickness \citep{zasov91,kregel05,sotnikova06,bm09,kh10}.
  In B17 we assumed that the disk mass is $M_d = 2 \pi \, \Sigma_0 \, h^2$, and the total mass of the galaxy within four disk
scale lengths is $M_t = 4\, V_{max}^2\, h / G$, 
$\Sigma_0$ is the central surface density of the disk,
$V_{max}$ is the circular velocity, and $G$ is the gravitational constant. 
For the disks in equilibrium in the vertical 
direction, their vertical scaleheight 
$z_0$ is \citep{spitzer42}: 
$\sigma_z^2 = \pi \, G \, \Sigma(R) \, z_0$, where 
$\sigma_z$ is the vertical stellar or gas velocity dispersion.

When combined with an assumption of the stellar disk marginal stability \citep{toomre64}, 
the radial stellar velocity dispersion $\sigma_R$ 
is $\sigma_R = 3.36\, G  \Sigma(R) / \kappa$, 
where $\kappa$ is the epicyclic frequency. If we take into account 
the non-axisymmetric perturbations and finite layer thickness, the 
minimum radial dispersion should be greater than 
$\sigma_R \geq Q 3.36\, G  \Sigma(R) / \kappa$, where $Q>1$ is the 
Toomre parameter
The value of $ Q \approx 1.4$ is 
consistent with the empirical criterion of gravitational stability 
\citep{kennicutt89}. 
The epicyclic frequency at the region of the flat
rotation curve is $\kappa = \sqrt{2} V_{max} / R$, and 
we get $z_0 \sim (M_d/M_t) h$. 
At $R = 2\,h$ the total-to-disk mass ratio is 
$M_t / M_d \, \gtrsim \, 1.2 \, (\sigma_z / \sigma_R)^2 \, 
(h / z_0) \, (Q / 1.4)^2$.

Note that we explore the case with thin disks and low vertical 
velocity dispersion, and refer to epochs when the galactic disks were mostly gaseous, for which case
$\sigma_R = \pi\, G\, Q\,  \Sigma(R) / \kappa$ \citep{safronov60} 
and $(\sigma_z / \sigma_R)$ = 1.
As it was further noticed by B17, the proximity of the 
vertical velocity dispersion in the stellar and gas disks, as well as the 
small disk thickness, can make the pure gas disk stability criterion 
equation applicable in the case of the superthin disks.
In this case $M_t / M_d \, \gtrsim \, 1.1\, (h/z_0)\, (Q/1.4)^2$.
We assume that the vertical velocity dispersion never falls 
below $\sigma_z  = 10 / \sqrt{3}$ km/s. 
For $\mu = M_t / M_d$ and $Q = 1.4$ B17 obtain 
\begin{equation}
\label{eq1}
h/z_0 \lesssim (81/\mu) \, V_{100}^2 \, ,
\end{equation}
where where $V_{100} = V_{\rm max} / 100$ \kms.

The relationship
in eq.(\ref{eq1}) corresponds to a marginally low vertical velocity dispersion in the stellar component of the order of 6 \kms, 
which is the same as in the star-forming gas component
\citep{kennicutt89,mogotsi16}. 

\begin{figure}
\epsscale{1.20}
\plotone{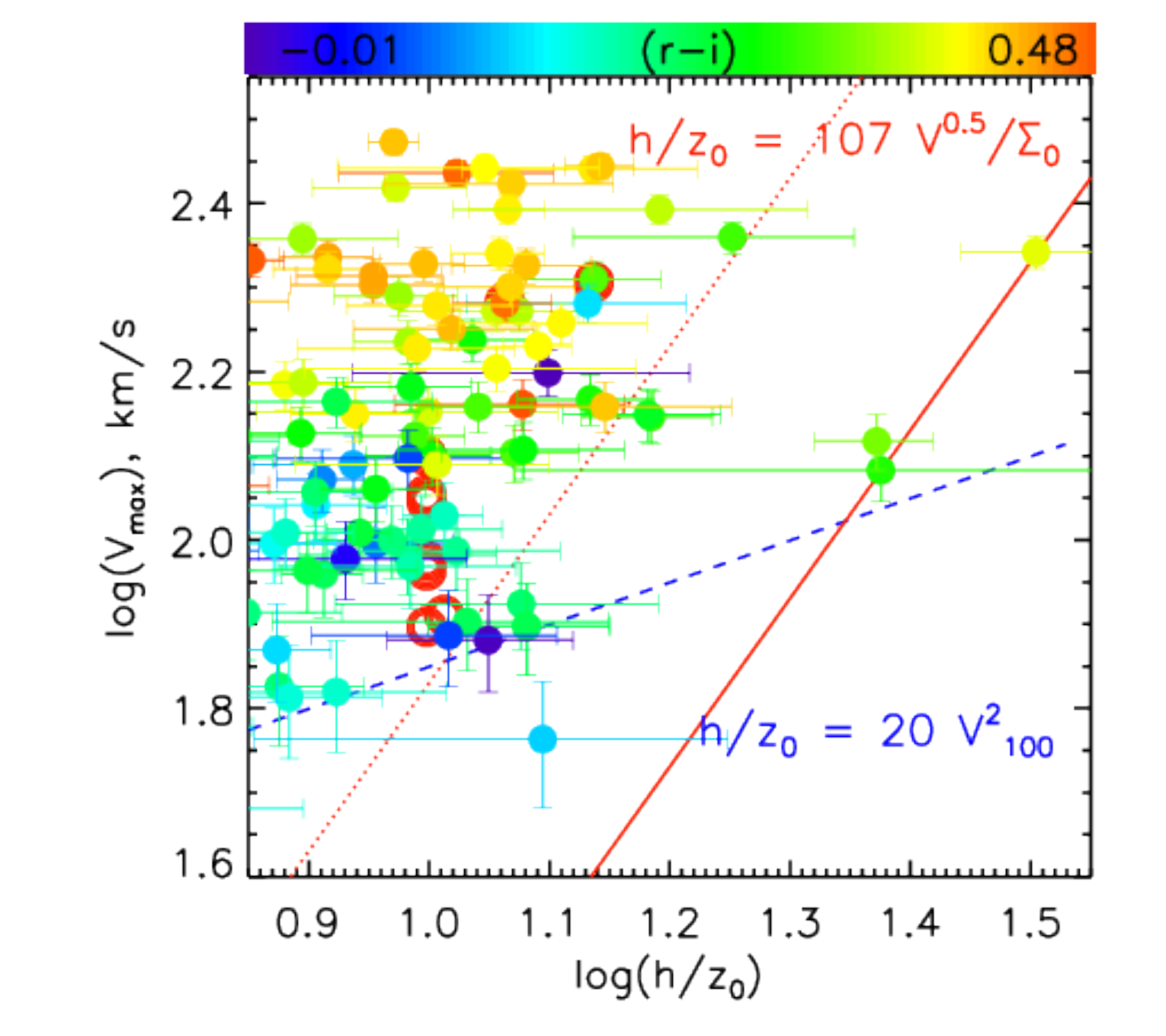}
\caption{
The rotation curve maximum \Vmax{} versus the scale ratio $h/z_0$. 
The colored bullets show our sample of galaxies. The symbol color indicates the galactic (r-i)$_0$ (upper bar).
The red open circles designate the data available from literature. 
The dotted red line marks the case of the disk surface density threshold found by B17.
The solid red line indicates our new density threshold of 49 \Msun{} found for our sample.
The dashed blue line is the case of the extremely small disk thickness regulated by eq.(\ref{eq1}). 
\label{fig4}}
\end{figure}

B17 noticed that the eq.(\ref{eq1}) does not limit the scale ratio $h/z_0$, and we could observe arbitrary thin galaxies under
reasonable assumptions on the $\mu$. The lack of observations of extremely thin galaxies could be explained by an additional constraint 
on the minimum stellar surface density $\Sigma_*$ that allows disks to form. In this case the \Vmax{} -- disk 
scale ratio relation looks like 
\begin{equation}
\label{eq2}
h/z_0 = 107 \, V_{100}^{0.5} / \Sigma_{0,c}
\end{equation}
(see B17).
The limited sample of galaxies considered by B17 suggests the minimum disk central surface density of 88 \Msunpc. 
Now we have collected a much larger sample of galaxies and can review this figure. Fig.~\ref{fig4} shows the 
$h/z_0$ -- \Vmax{} diagram for our galaxies. The red dotted line corresponds to the former value of $\Sigma_{0,c}$ = 88 \Msunpc.

Fig.~\ref{fig4} demonstrates a lower threshold on the limiting central surface density $\Sigma_{0,c}$. 
It needs to be at most
49 \Msunpc{} to explain the most rightward points on the diagram. 
At the same time, most of the points meet the threshold set by eq.(\ref{eq2}).
The outlier below the blue dashed line can be explained by a lower than assumed
spherical-to-disk mass ratio $\mu$. We purposely show the galaxies with the $h/z_0$ scale
ratio below 9 to indicate the trends created by the galaxies on the border of the superthin criterion. 
Those galaxies were selected among the most blue and visually thin galaxies, without the formal, numerical
selection by their thickness.

\section{Simultaneous Rotation Curve and disk Thickness Modeling}

\subsection{Assumptions for the Modeling}

The rotation curve modeling demonstrates a large variability in the obtained parameters of 
stellar disks and dark halos between the ”maximum disk” and ”maximum halo” cases
\citep[e.g.][]{vanalbada86}. Introduction of an information about the vertical velocity 
dispersion helps constrain the modeling significantly \citep{angus15}. 
In the case of edge-on  galaxies, using the disk thickness
allows us to constrain the rotation curve modeling \citep[e.g.][]{kh10}. 
Note that in a bulgeless galaxy with one exponential stellar disk with known scale length 
and a spherical dark matter halo only, the additional constraint from the stellar disk 
thickness leads to only two free parameters
in the modeling: the scale length of the halo and the central surface brightness of the disk. 
The latter also can be constrained by known central surface brightness and an estimate of the mass-to-light ratio
from the galactic color. The dust extinction makes such an estimate very uncertain in 
edge-on galaxies, and the stellar surface densities estimated from an edge-on galaxy decomposition should be used in practice only as a lower limit. 
As we mention in \S3.1, our galaxies have significantly different
scale lengths in their old and young stellar populations traced by the optical and NIR photometry, 
which can be interpreted as a presence of 
at least two stellar disks with different scale lengths but similar scale heights. 
This implies complications that we intend to introduce to the simple single disk-in-halo model. 

Similar to \S\ref{vrothz} and B17, we assume that the volume density distribution $\rho_d$ in the stellar disk follows exponential
law along the radius $r$ and isothermal along the vertical direction $z$: 
\begin{equation}
\rho_d(r,z) \,=\, \rho_{d,0}\, exp(-r/h)\, sech^2(z/z_0) .
\label{rhodisk}
\end{equation}
The dark matter halo is assumed to be spherical and pseudo-isothermal with 
its density 
\begin{equation}
\rho_{halo}(R) = \rho_{halo,0} [1 + (R/a_{halo})^2]^{-1}   , 
\label{rhohalo}
\end{equation}
where the $\rho_{halo,0}$  is the central density, 
$a_{halo}$ is the dark matter distribution scale length, and R is the three-dimensional distance to the center $R^2=r^2+z^2$. 
In this paper we include the stellar disk thickness into
the analysis via the relation between the inverse relative disk thickness $h/z_0$ and the spherical-to-disk
mass ratio shown by \citet{kh10}. The latter can be parameterized as 
\begin{equation}
M_{sph}/M_{disk} = -1.256 + 0.399\, (h/z_0).
\label{MhMdhz}
\end{equation}

The gravitational potential of an exponential disk contributes to the rotation curve as
\begin{equation}
V_d^2(R) \,=\, 4 \pi G \Sigma_0 h y^2 [I_0(y) K_0(y) - I_1(y) K_1(y)]   ~,
\label{v2disk}
\end{equation}
where y=0.5R/h, $\Sigma_0$ is the disk central surface density, 
$I_n(y)$ and $K_n(y)$ are the modified Bessell functions of the first and second kind
\citep{bt87,jimenez03}.

The contribution of the pseudo-isothermal spherical dark halo is assumed to be
\begin{equation}
V_{halo}^2(R) \,=\, 4 \pi G \rho_0 a_{halo}^2 [1 \,-\, ( (a_{halo}/R) ~arctan(R/a_{halo}))] 
\label{v2halo}
\end{equation}

We anticipate to find an exponential pseudo-bulge in the thin, star forming galaxies, so in rare cases when \citet{EGIS}
suggested a bulge in our galaxy, we described its contribution $V^2_b$ similar to equations (\ref{rhodisk},\ref{v2disk}) with 
its own bulge scale length $R_b$.

The gas contribution to the overall gravitational potential is small in large galaxies like the Milky Way.
In small, LSB galaxies its fraction can be noticeable
\citep{vdKruit01}, and we cannot ignore it in the case of our superthin
sample. However, direct measurements of the neutral gas distribution across the disks were performed for just a few,
nearby galaxies 
\citep{uson03,matthews08,obrien10}.
Fortunately, our galaxies are either large and
dynamically evolved, or small and under-evolved (see B17). In the former case we can assume that
the galactic gas contribution to the overall gravitational potential is very small. In the latter case we 
expect to see significant gas contribution, but the gas galactic subsystem resembles the young stellar 
population by the structural parameters. In this case we can refer to the young stellar disk as to a
mix of gas and stars, with the volume density parameterized by equation (\ref{rhodisk}) and the rotation curve
contribution described by equation (\ref{v2disk}). 

The resulting rotation curve $V_{all}$ is the sum of $V_d$, $V_{halo}$
and $V_b$ (if detected) in quadrature $V_{all}^2 = V_d^2 + V_{halo}^2 + V_b^2$. We add the asymmetric drift correction \citet[e.g. in][]{bt87} obtained for the case of gas to the velocity above in quadrature. 

Since all our galaxies are seen edge-on, we take into account the projection effects by integrating the $V_{all}$
along the line of sight without accounting for a small deviation from the perfect edge-on inclination. 
In this case we assume that the observable model velocity $V_{mod}$ is
\begin{equation}
V_{mod}(x) \,=\, \int_{los} V_{all}(r) \, \rho_g(l) dl ~ / \int_{los} \rho_g(l) dl
\label{vint}
\end{equation}
Here $x$ is the distance from the center along the major axis, $\rho_g$ is the emitting gas density, which
we assume to be distributed similar to $\rho_d$, and $l$ is the emitting element depth along the line of sight, $l^2=r^2 - x^2$.
The depth $l$ is normalized such that $l=0$ at the near edge of the disk. 

\citet{matthews01,maclachlan11} came to a conclusion that small 
and thin LSB galaxies have low overall dust extinction. Their face-on optical depth is much less than unity. 
The dust scale height in those galaxies is comparable to their stellar scale height. 
We do not neglect the dust 
extinction, given its contribution in regular galaxies can significantly change the shape of rotation 
curve in the inner disk regions \citep{bosma92,zasov03}. We assume that the dust forms an exponential
disk and that the dust extinction coefficient \citep{xilouris99}  can be described as
\begin{equation}
\kappa (r,z) \,=\, \kappa_0\, exp(-r/h_{dust})\, sech^2(z/z_{dust}) ,
\label{rhodust}
\end{equation}
where $\kappa_0$ is the central value. 
The resulting dust extinction $\tau \,=\, \int_{los} \kappa \, dl$.
We expect that the $h_{dust}$ can vary between $h$ and 2$h$,  
and the $z_{dust}$ vary between 0.5$z_0$ and $z_0$
\citep{xilouris99,matthews99,yoachim06,bianchi07}.
We note in advance that the resulting extinction is found to be small, and we cannot notice significant difference between
the cases of $h$ and 2$h$ for the dust scale length and between 0.5$z_0$ and $z_0$ for the dust scale height
in the modeling described below in the paper.
Also, the reasonable range of the dust extinction values in equation (\ref{rhodust}) that brings up the $\tau_V$ at the center of the disk integrated along the vertical direction (i.e. its face-on value)
in the $V$-band between 0 and 1 does not affect the resulting modeled rotation curves significantly. 
For certainty, we assume that $h_{dust} = h$ and $z_{dust} = z_0$ and the central face-on $\tau_V = 0.5$ in the 
models described below, unless the dust extinction parameters were allowed to vary as free parameters (Models 6 and 7).
We neglect the light scattering because it makes only a small effect in the detailed radial transfer simulations on
edge-on galaxies \citep{bianchi07}. 
The dust extinction $\tau$ is added to equation (\ref{vint}).
The finite FWHM for the spectral observations is taken into account.
The model rotation curve is convolved with a Gaussian kernel that corresponds
to the reported seeing (see Table~\ref{tab:tab2}), in order to
obtain the final rotation curve before the comparison with the observable one.

\subsection{The Model Constraints and Free Parameters}

The set of equations (5-11) allows us to build model rotation curves $V_{mod}$ and compare them to our observations $V_{obs}$. 
We employ additional constraints for the 
vertical structure information that
helps mitigate ambiguities in the rotation curve modeling. We develop several models with different sets of parameters and constraints from observations.
The variety of models helps us understand how different assumptions affect the model output.
The model parameters were evaluated via the chi-square optimization. We minimized the sum
$\Sigma \, (V_{mod} - V_{obs})^2 \, / \, \sigma_{obs}$, where the $\sigma_{obs}$ designates uncertainties of the 
rotation curve points. 

We incorporate the optical and NIR structural parameters
from Table ~\ref{tab:tab3}.
We indirectly use the stellar densities estimated
from the optical \citep{EGIS} and NIR \citep{nir}
central surface densities via the mass-to-light values tabulated by \citet{bell03}. 
The estimated densities set the lower limits, while the real density values should be higher due to the extinction corrections and the gas contribution. 

\noindent {\it Model 0}\\
This is an over-simplified, toy model that assumes only one disk whose scale length and height correspond
to the optical observations. It has two free parameters (in the case of no bulge): the dark halo scale
length $a_{\rm halo}$  and the disk central surface density. While the model does not take into account additional constraints 
that come from the optical or NIR observations, it helps establish fiducial chi-square values for each galaxy, which
can be used for comparison with the other models. 

\noindent {\it Model 1}\\
Here we assume the presence of two disks -- young and old. The young disk has the $h$ and $z_0$ estimated
from optical SDSS images, while the old one has the scales estimated from the NIR data. We have two surface density parameters
instead of one, so the number of free model parameters increases by one with respect to the Model 0.

\noindent {\it Model 2}\\
This model has the same parameters as the Model 1, but the halo-to-disk mass ratio does not follow the 
eq.(\ref{MhMdhz}). Instead, this is a free parameter of the model.

\noindent {\it Model 3}\\
In this case we assume a single stellar disk, but its scale length is unknown and is a free parameter. 
The unknown scale length $h_d$ varies between the `young' $h$ and `old' $h_{nir}$ values. 
The single disk surface density is also a free parameter. 
The assumptions can be interpreted as the presence of an infinite number of disks, 
whose `effective' scale length $h_d$ and total central surface brightness are evaluated in the model.

\noindent {\it Model 4}\\
This model is the same as Model 2, but the central surface densities of the old and young stellar disks have additional 
constraints. They are limited by the minimum surface densities estimated from the optical and NIR photometry.
The upper limits are 2 magnitudes higher than the lower limits, which accounts for a possible dust extinction
and for the gas addition. 

\noindent {\it Model 5}\\
Here we modified the Model 3 by adding constraints on the central surface density of the disk the same way as
for the Model 4, where the minimum surface densities were estimated from the photometry. 

\noindent {\it Model 6}\\
We modified Model 5 by adding the central dust absorption $\kappa_0$ from equation (\ref{rhodust}) as a free parameter,
while all other parameters and constraints were kept unchanged. 

\noindent {\it Model 7}\\\
Here we added the central dust absorption $\kappa_0$ as a free parameter to the Model 0.

The parameters of the models are summarized in Table~\ref{modelpar}. 
Note that we used one additional free parameter -- the bulge scale -- 
for the galaxies with bulges. The mass of the bulge was estimated by assuming the bulge-to-disk mass ratio follows
the bulge-to-disk luminosity ratio from the EGIS catalogue.

\begin{table}
\centering
\setcounter{table}{4}
\caption{Parameters of the models}
\begin{tabular}{llr} 
\hline
Model &  Free parameters & Constraints \\
\hline
Model 0 & $\Sigma_0$, $a_{halo}$                                                                         & None \\
Model 1 & $\Sigma_{0,young}$, $\Sigma_{0,old}$, $a_{halo}$                              & None \\
Model 2 & $\Sigma_{0,young}$, $\Sigma_{0,old}$, $a_{halo}$, $M_{halo}/M_d$  & None \\
Model 3 & $\Sigma_0$, $a_{halo}$, $h_d$                                                             & $h_d$    \\
Model 4 & $\Sigma_{0,young}$, $\Sigma_{0,old}$, $a_{halo}$, $M_{halo}/M_d$  & $\Sigma_{0,young}$, \\
        &   & $\Sigma_{0,old}$ \\
Model 5 & $\Sigma_0$, $a_{halo}$,  $h_d$                                                            & $\Sigma_0$, $h_d$ \\
\hline
Model 6 & $\Sigma_0$, $a_{halo}$, $h_d$ , $\kappa_0$                                       & $\Sigma_0$, $h_d$ \\
Model 7 & $\Sigma_0$, $a_{halo}$, $\kappa_0$                                                    & $\Sigma_0$, $h_d$ \\
\hline
\end{tabular}
\label{modelpar}
\end{table}

\subsection{Results of the Modeling}

\begin{figure}
\epsscale{1.20}
\plotone{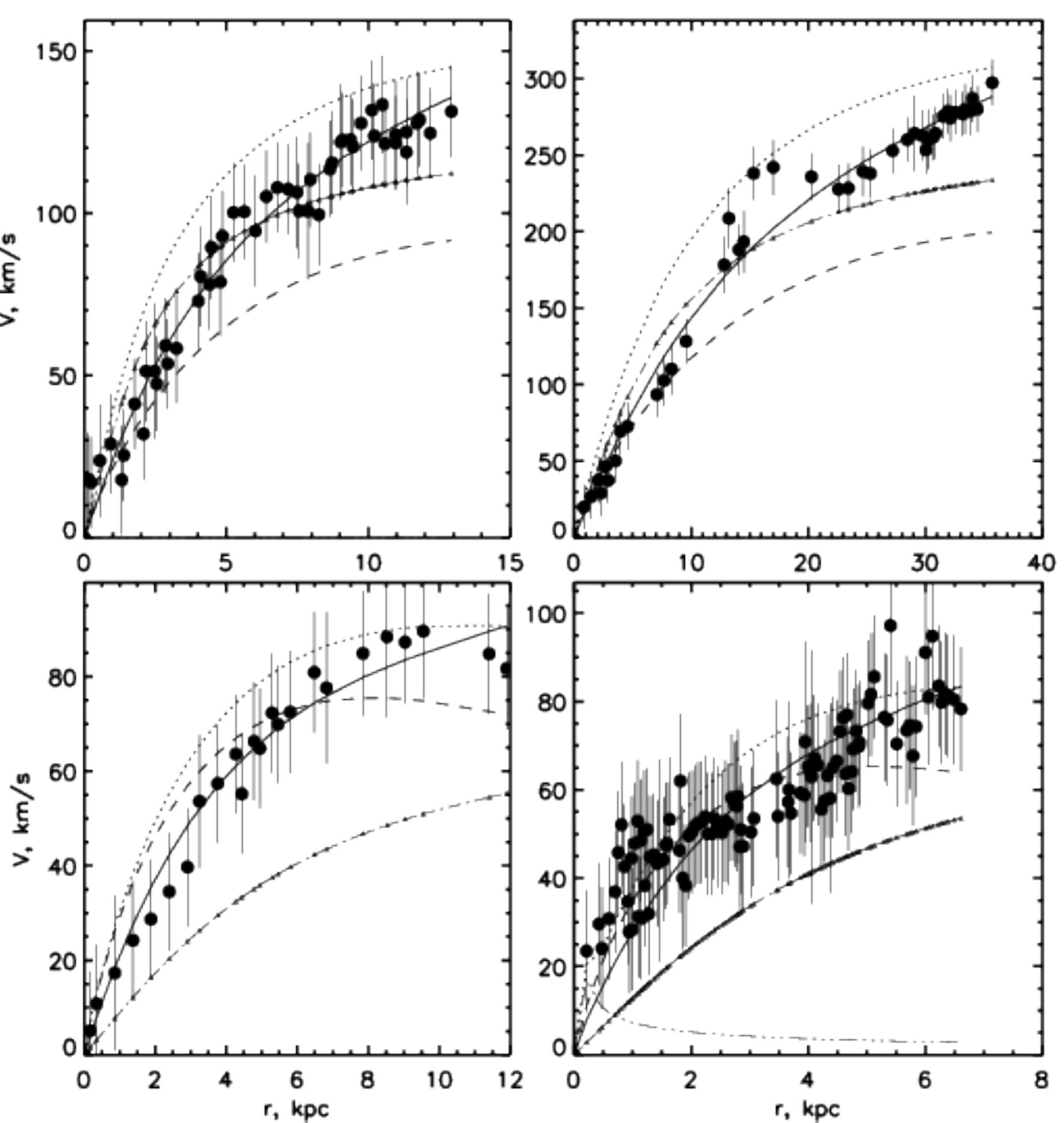}
\caption{Examples of modeling for some types of our galaxies in the frames of the Model 6.
The upper left panel demonstrates the modeling for a typical galaxy, EON\_19.768\_$-$0.139.
The dash-dotted and dashed curves designate the contribution from the dark matter halo and stellar disk, respectively. 
The dotted curve is the true rotation curve from all galactic components. The solid curve is the 
true rotation curve after the edge-on projection. The bullets with error bars represent the observing rotation curve. 
The upper right panel shows the same curves for a red superthin galaxy, EON\_149.150\_20.646.
The lower left panel is for a blue galaxy EON\_51.832\_3.920. The lower right panel shows
the modeling for a galaxy with small bulge, EON\_123.495\_45.742. The bulge contribution 
is shown with the dash-doubledotted curve. 
\label{fig5}}
\end{figure}

Fig.~\ref{fig5} shows examples of the chi-square minimization for some galaxies in our sample:
a red one (upper right panel), a blue one (lower left case), a galaxy with bulge (lower
right), and a galaxy in-between the red and blue superthins (upper left panel).
The curves designate the contributions of disk (dashed curve) and dark halo (dash-dotted curve)
to the rotation. The dotted curves show the total rotation curves from all galactic components. 
The solid curves show the results of the projection, i.e. the model observable rotation curves. 
The bullets with bars denote the observing data points and their uncertainties. 

\begin{figure}
\epsscale{1.00}
\plotone{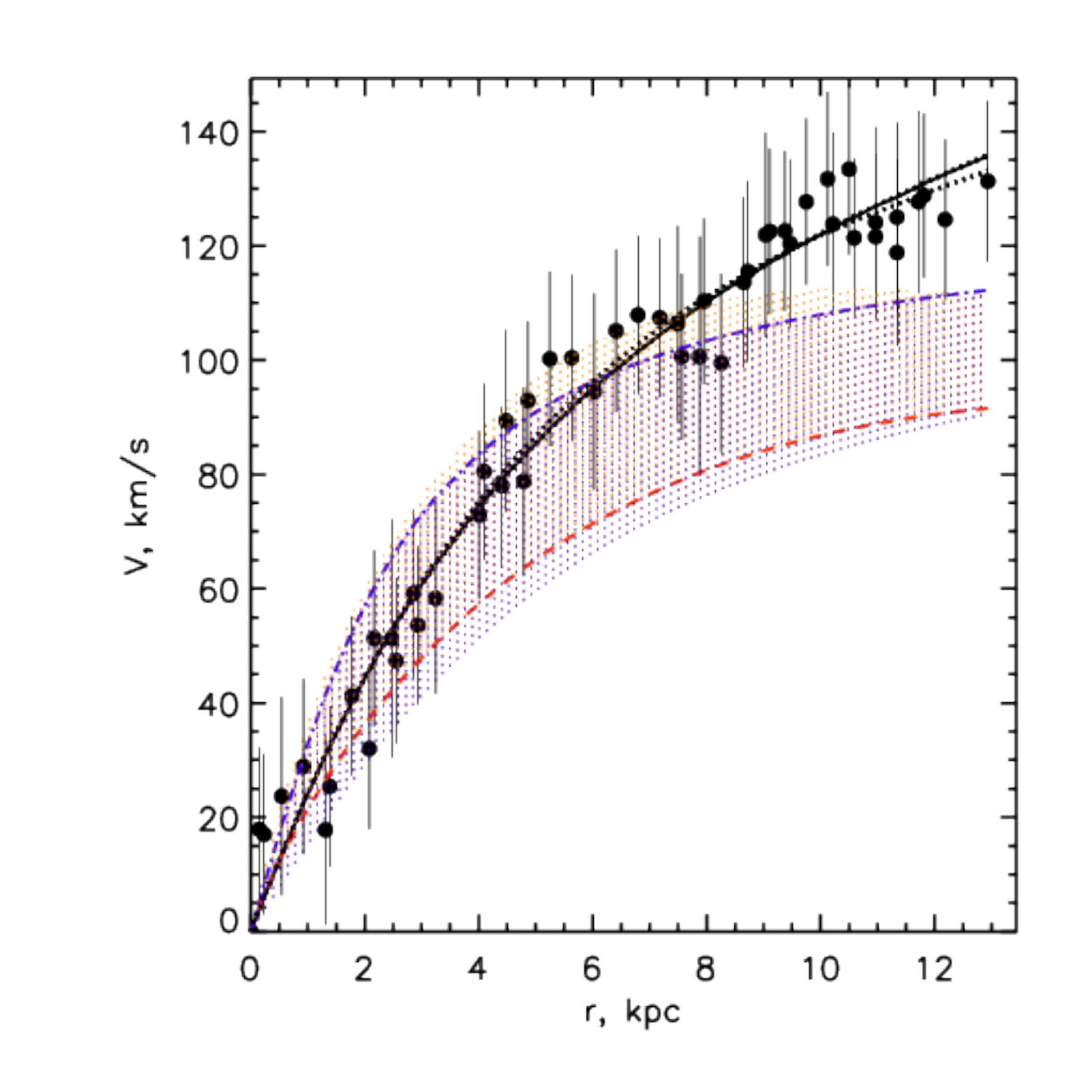}
\caption{The ranges of the disk (red) and halo (blue) rotation curves between different models
0-7 are designated with the shaded areas for the galaxy EON\_19.768\_$-$0.139. The
red dashed curve designates the disk-only rotation curve for Model 6, and the blue
dash-dotted line is for the halo in the same model. The black solid curve denotes the sum
rotation curve after the edge-on projection for the Model 6. All other models are 
designated with the dotted black curves. The bullets with error bars show the 
observing rotation curve. 
\label{fig6}}
\end{figure}

The modeling allows us to obtain parameters of stellar disks and dark halos for all our galaxies.
The parameters are shown in Table \ref{tab6}.
Fig.~\ref{fig6} demonstrates the resulting rotation curves
for different models (the solid and dotted black lines) and
the ranges of the rotation curves for the disk and halo 
components (shown with the shaded red and blue areas, respectively.)
Fig.~\ref{fig6} also shows an example of similar resulting rotation
curves after the edge-on projection obtained in different models 
while the disk and halo rotation curves can be significantly different.
On average, the models improve best-fitting chi-square values
with respect to Model 0. The least improving are Models 1 and 7 - 
their median chi-square difference from the Model 0 normalized by the latter
is about 2\%, whereas the most improving are the Models 5 and 6
(about 50\% of the median chi-square improvement.)
We select the Model 6 for the further analysis because it has 
the best overall chi-square values for the models, on average.  

The most consistent results between the models are obtained for the stellar disk scale length $h$ (when allowed to vary)
and the dark halo mass $log M_h$. The median value of the $h$ is greater by up to 3\% in  
models 0, 1, 2, 4, \& 7, or lower by 1\% in model 3, with respect to model 6. The median
value of $log M_h$ is less than 1\% in models 0, 1, 3, \& 7, and greater than 1\% in models 2,4, \&5
with respect to model 6. The disk central surface density is not farther than 10\% from model 6's
values in models 0, 3, 5, \& 7. This value differs from model 6 by more than 50\% in model 4.
The halo scale $a_{halo}$ is within 10\% from that in model 6 in all models except
for models 2 and 4, where they are longer by up to 84\%. We notice that the halo scale shows a large overall scatter between the models. 
The halo-to-disk mass ratio is 30\% lower 
in models 0, 1, \&6, very consistent in models 3 \& 5, and is greater by 140\% in model 4,
with respect to model 6. Not a surprise that the parameters closely related to measurable
values - $h$ and $log M_h$, are estimated the most reliably and therefore are consistent 
from model to model, while the dark halo structure parameter
$a_{halo}$ varies significantly between our different models.

Having information about the gravitational potential of all galactic components and the 
surface density, we can assess the dynamical state of the disk as in \citet{spitzer42}
by estimating its vertical
velocity dispersion at 2$h$ distance from the center as
\begin{equation}
\sigma_*(2h) \,=\, [\pi G \Sigma_0 e^{-2} z_0 ]^{1/2}     ~~.
\label{sigma2h}
\end{equation}

\begin{figure}
\epsscale{1.00}
\plotone{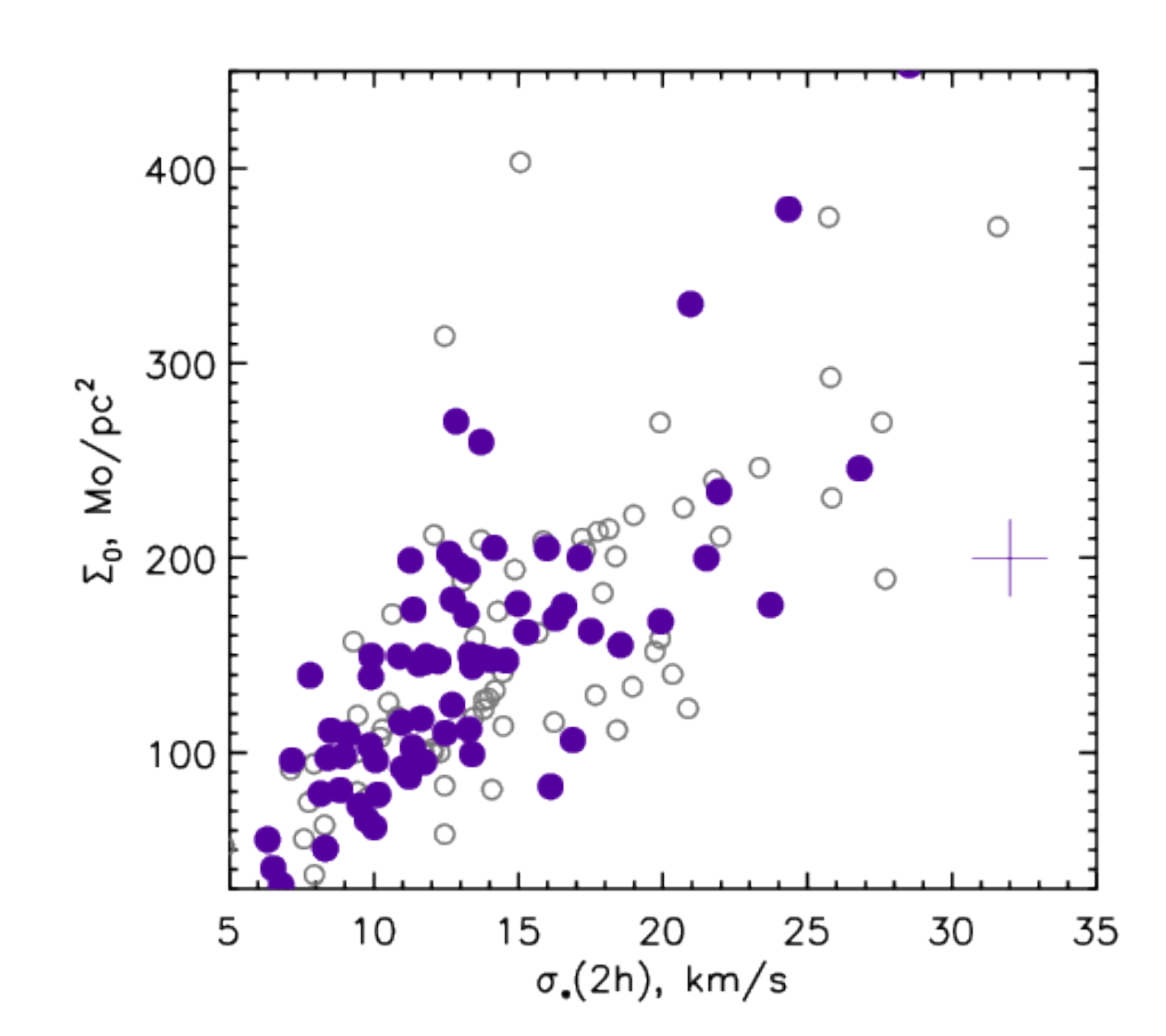}
\caption{
The disk $\Sigma_0$ - $\sigma_*(2h)$ diagram. The vertical velocity dispersion at two disk scales from 
the center $\sigma_*(2h)$ is estimated via equation (\ref{sigma2h}). The grey circles represent the worse than average
fitting (chi-squared is greater than the median of the sample), while the blue bullets designate our galaxies with good fitting.
The blue cross demonstrates the typical error bars along the axes.
\label{fig7}}
\end{figure}

Fig.~\ref{fig7} shows a relation between the central surface density and the 
vertical velocity dispersion estimated at two disk scale lengths. 
We check if the quality of fitting, in terms of the chi-square value, affects
the results and the dependencies between the output parameters via dividing our 
objects by better- and worse-fitted samples. We divide the sample using the median 
chi-square value, and show results with different symbols in Fig.~\ref{fig7}
and further figures in the paper. As it can be seen in Fig.~\ref{fig7}, both 
better- and worse-fitted galaxies show the same trends, and our conclusions 
do not depend on the quality of fitting. Same conclusion can be inferred for
the further figures.

We calculate the errors of resulting model output parameters for a typical 
galaxy EON\_19.768\_--0.139 shown in Fig.~\ref{fig5} via running the model 6
for it one hundred times with slightly changed input parameters. Each run we allow the input model parameters - stellar disk scales and velocities - to take random
values according to their uncertainties from observations. We assume 
that the input errors are distributed normally. As a result, the modeling
is run with slightly different parameters, and therefore produces different
results. The typical uncertainties of resulting parameters are found
as standard deviations of the output values.

Note that equation (\ref{sigma2h}) suggests a correlation between the $\Sigma_0$ and $\sigma_*(2h)$.
We demonstrate Fig.~\ref{fig7} to show that the lowest surface density galaxies in our sample 
with $\Sigma_0 \,<$ 120 \Msunpc\, have very low velocity dispersion under 10 \kms{}.
It confirms that the thinnest disks have the lowest stellar densities, 
which also suggests that their stellar disks did not experience
significant dynamical heating during the history of the galaxies, 
and kept their vertical dispersion close to that of the galactic gas
through the galactic history.

\begin{figure}
\epsscale{1.20}
\plotone{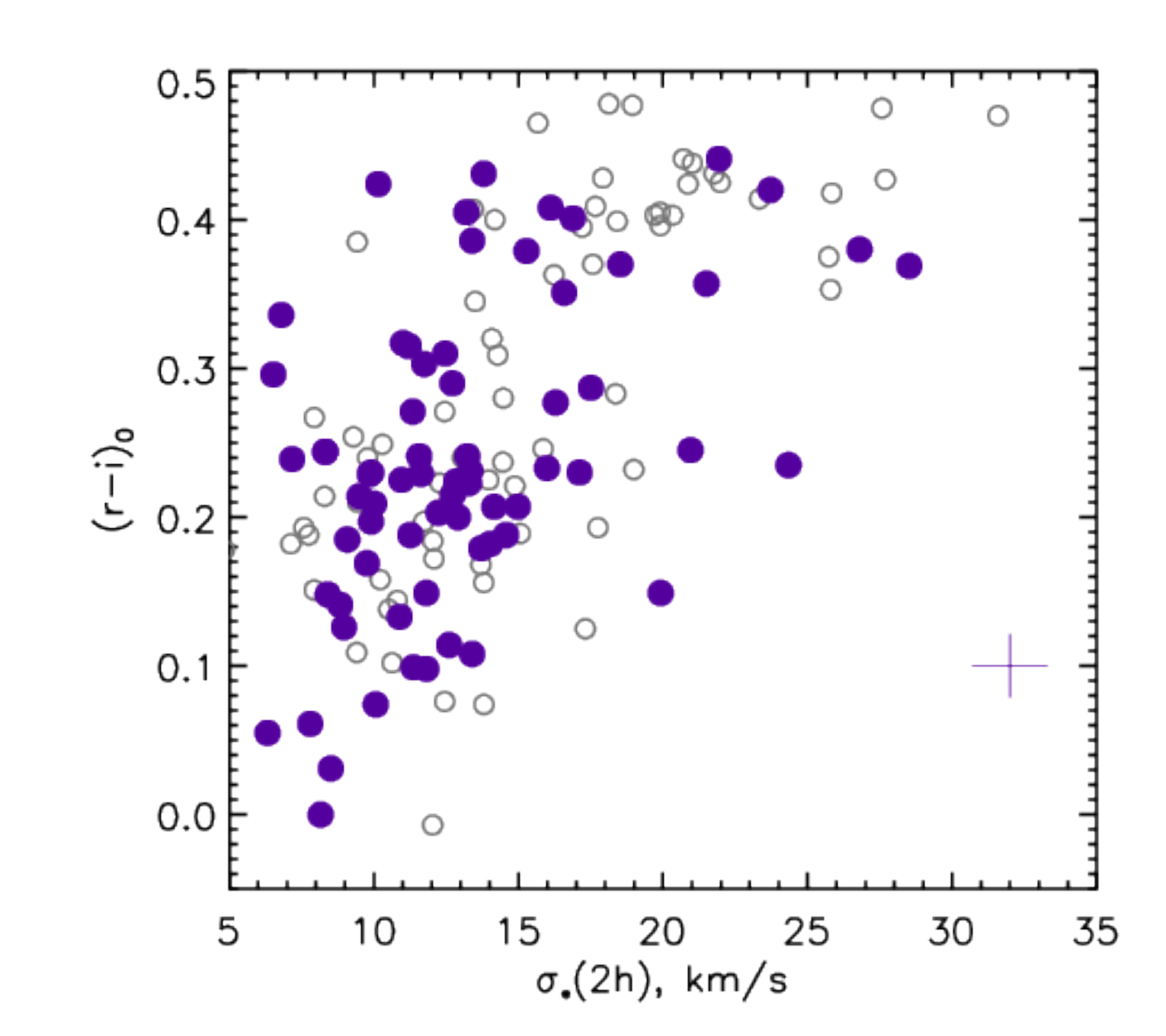}
\caption{
The color -- stellar velocity dispersion diagram. The designations are kept as in Fig.~\ref{fig7}.
\label{fig8}}
\end{figure}

If we replace the surface density with galactic colors as in Fig.~\ref{fig8},
it is clear that the most dynamically under-evolved galaxies are among the blue subsample with the 
colors $(r-i)_0 < 0.24$~mag. It is also interesting to notice that the vertical velocity dispersion 
falls below 10 \kms{} in the galaxies whose disk central surface density goes
below the average dark halo surface density (90--110 \Msunpc) inferred by \citet{kormendy04,dipaolo19},
i.e. the dark halo dominates in them everywhere by the surface density. 
On a contrary, the reddest galaxies in the sample indicate rather high vertical velocity dispersion
with respect to the gas, a few tens \kms.

\begin{figure}
\epsscale{1.00}
\plotone{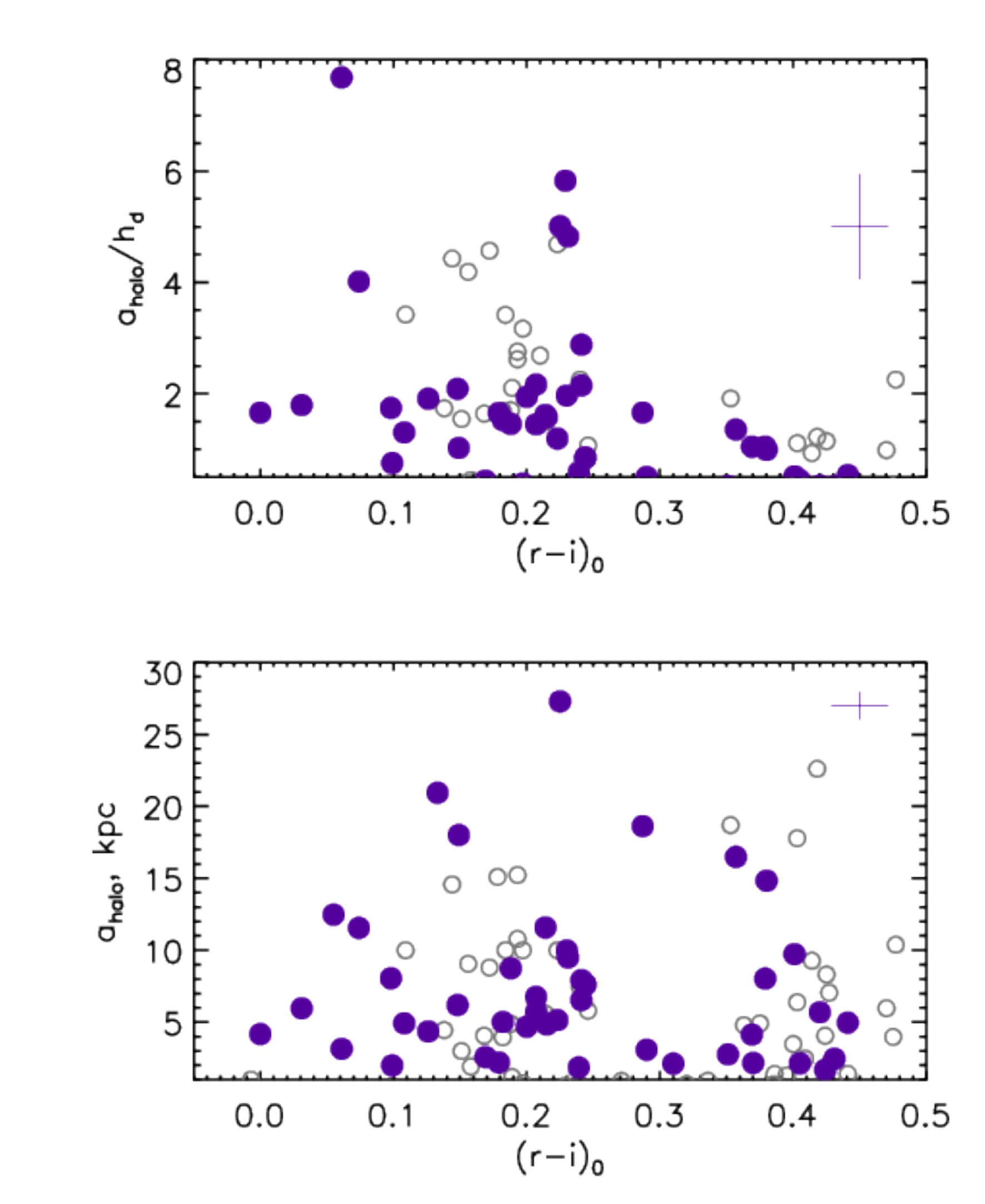}
\caption{
Upper panel: the dark halo scale normalized by the disk scale length $a_{\rm halo} / h_d$ versus the galactic color $(r-i)_0$.
Lower panel: the dark halo scale $a_{\rm halo}$ versus the galactic color $(r-i)_0$.
The designations are kept as in Fig.~\ref{fig7}.
\label{fig9}}
\end{figure}

Although the dark matter halo parameters -- mass and scale -- have different absolute values between the models,
we can notice a common behavior of these parameters considered in the relative sense. 
We observe the following common tendencies. 
While a correlation between the dark halo mass $M_h$ and the maximum rotational velocity is expected,
the halo scale length $a_{\rm halo}$ shows the inverse dependence of the \Vmax, and also of
the color, see Fig.~\ref{fig9}. 
It is notable to mention that the blue galaxies tend to have extended
dark halos, and the halo-to-disk scale ratios are high for the blue STGs as well. 
On a contrary, 
the red STGs have rather short-scaled dark halos (small $a_{halo} / h_d$ ratios), 
which is in agreement with results of numerical simulations by \citet{dipaolo19}. 

\begin{figure}
\epsscale{1.00}
\plotone{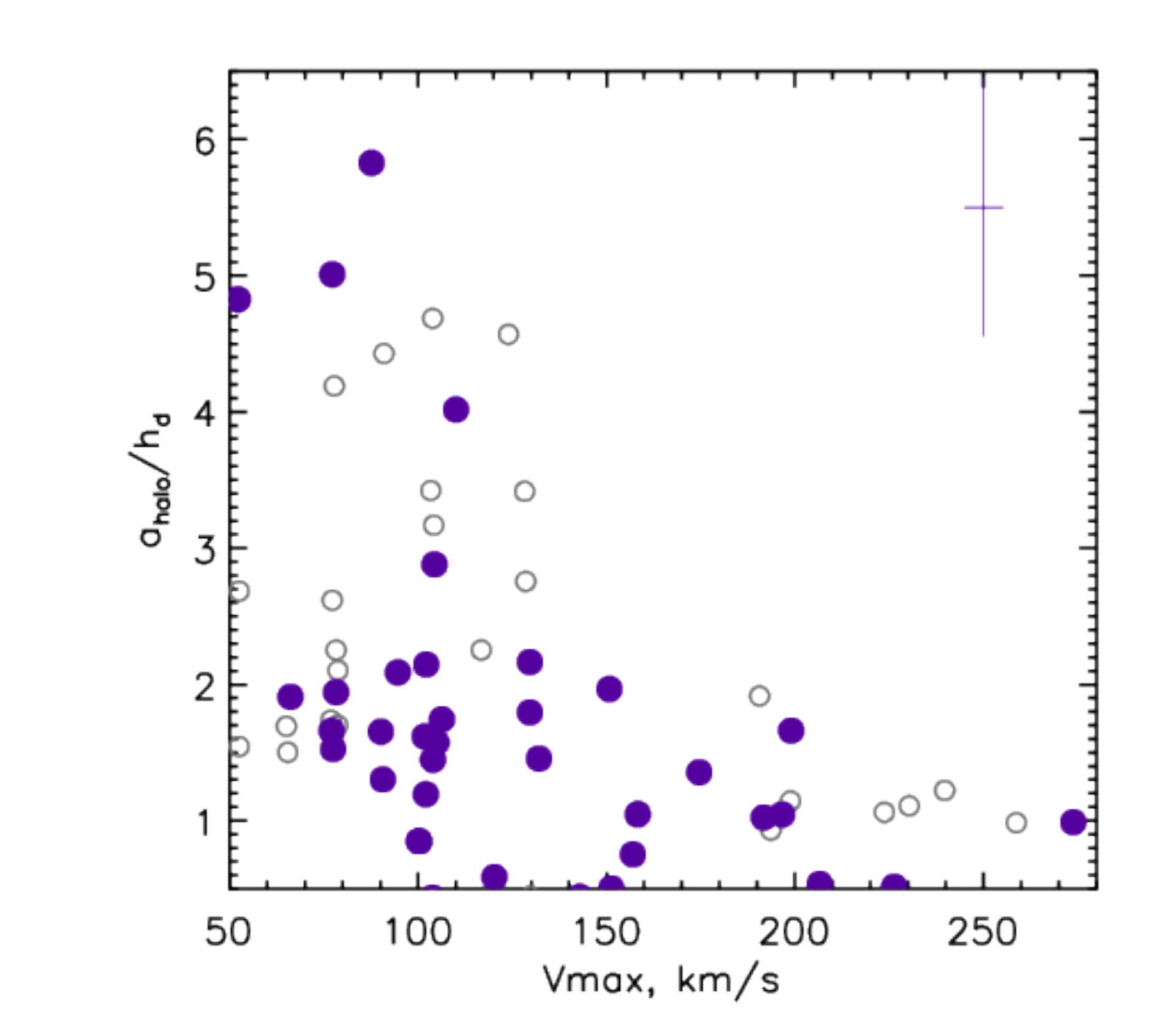}
\caption{
The dark halo scale $a_{halo}$ -- the maximum rotation velocity \Vmax{} diagram.
The designations are kept the same as in Fig.~\ref{fig7}.
\label{fig10}}
\end{figure}

The galaxies with high \Vmax{} also have compact halos, whereas the low rotation curve maxima
follow the rarefied halos (Fig.~\ref{fig10}) in a combination with LSB disks, as in simulations by \citet{maccio07}.

\begin{figure}
\epsscale{1.00}
\plotone{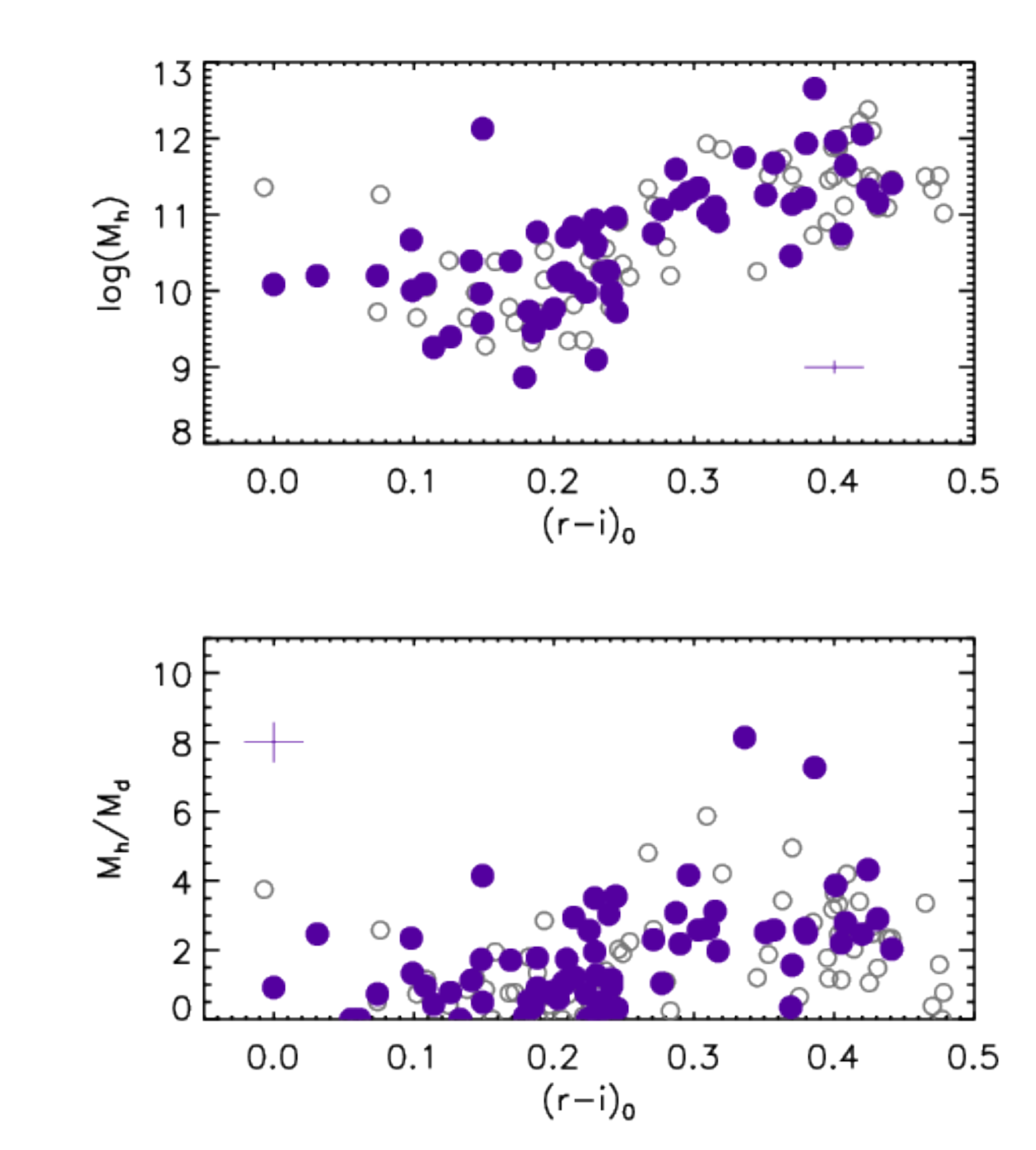}
\caption{
The M\_halo - (r-i) and  M\_halo / Md- (r-i) panels.
The designations are kept the same as in Fig.~\ref{fig7}.
\label{fig11}}
\end{figure}

The masses of dark halos correlate with the colors $(r-i)_0$, and the most massive halos are observed in the
reddest STGs in our sample, see Fig.~\ref{fig11}. 
The halo-to-disk mass ratio, which reveals the dark-to-light mass in the galaxies,
does not show a clear systematic variation with the galactic color (Fig.~\ref{fig11}) or with \Vmax.

\subsection{Modeling with the Observed Disk Thickness Radial Distribution}
A way to include the observed radial distribution of the stellar disk thickness into the modeling was presented by \citet[e.g.][]{narayan02,banerjee08,banerjee17}. We 
follow it, and start with the Poisson equation for the thin disk:
\begin{equation}
\frac{\partial^2\, \Psi_s}{\partial z^2} +
\frac{\partial^2\, \Psi_g}{\partial z^2} +
\frac{\partial^2\, \Psi_h}{\partial z^2} = 
4\pi G (\rho_s + \rho_g + \rho_h),
\label{eq:poisson}
\end{equation}
where the subscripts "s", "g", and "h" refer to the stars, gas, and halo, respectively. 
We combine equation~(\ref{eq:poisson}) with the hydrostatic equilibrium equation in the vertical direction
\begin{equation}
\frac{ <(c_z)^2_i>}{\rho_i}
\frac{\partial \rho_i}{\partial z} = K_s + K_g + K_h ,
\label{eq:hydro}
\end{equation}
where $K_i = \partial \Psi_i / \partial z $, and the "$i$" subscript refers to either stars or gas or dark halo, and $<(c_z)^2>$ is the root mean square of the vertical velocity. 
By combining equations (\ref{eq:poisson}) and (\ref{eq:hydro}) we obtain a system of equations for each component. We assume that the density distributions in the halo and disks follow equations (\ref{rhodisk}, \ref{rhodust}, \ref{rhohalo}) and use equations for the gravitational potentials by \citet{vdkruit82b,narayan02,vdkruit88}.
Since we intend to compare the resulting vertical disk scale with observed $z_0$ estimated for the case described by equation (\ref{rhodisk}), we substitute the functional form to equations (\ref{eq:poisson}) and (\ref{eq:hydro}) and solve them on iterative way. Given the same functional forms of the galactic components, the rotation curve fitting was kept identical to \S4.1.

\begin{figure}
\epsscale{1.00}
\plotone{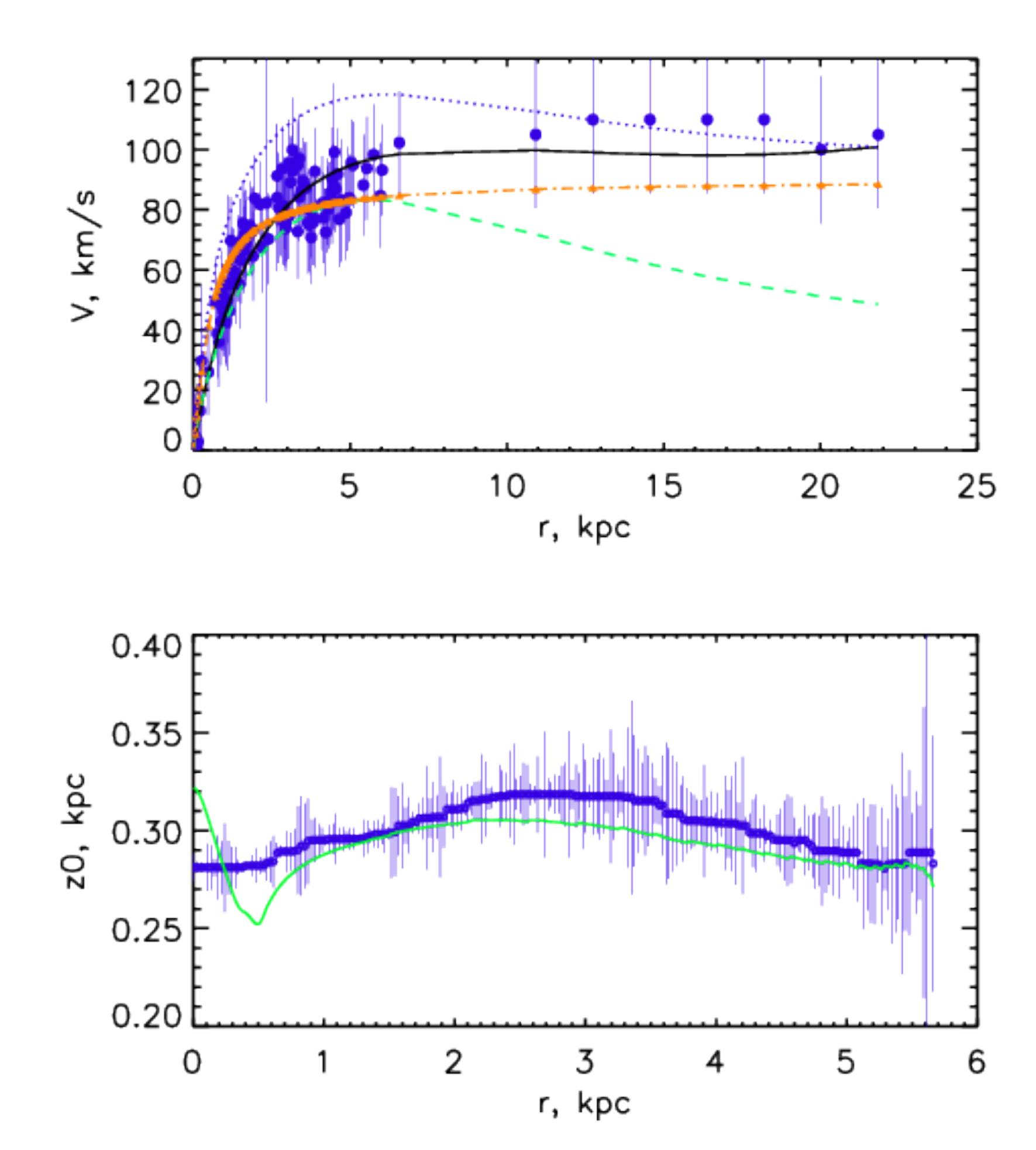}
\caption{
Results of simultaneous modeling of the rotation curve
and the disk thickness radial distribution for the galaxy 
UGC~7321. The upper panel has the same designations as in 
Figure \ref{fig5}. Our ionized gas rotation curve was updated with the 
rotation measurements from \citet{uson03} in the outer disk, beyond 7 kpc 
from the center. The lower panel shows the observed (blue dots with 
error bars) and model (green curve) disk thickness radial distribution.
\label{model_v_z}}
\end{figure}

The obtained model disk thickness $z_0$ is integrated along the line of sight in the edge-on, dusty disk with equations analogous to (\ref{rhodust}) and (\ref{vint}). The resulting model disk thickness is compared to the observed ones along the radius in same manner as it was done to the velocities. We find the best-fit parameters by minimizing the chi-square of differences between observed and model parameters. The minimized chi-square is the sum of chi-squares obtained for the velocity and disk thickness separately. Note that the scale height distribution provides twice as many data points as the rotation curve for the minimization. To equalize contributions of the two factors to the overall chi-square, we attempted to multiply the chi-squares by the number of data points, as well as by the squared number of data points. The resulting best-fit parameters did not change significantly. 
We perform this modeling for the galaxy UGC~7321. Figure \ref{model_v_z} shows results of simultaneous fitting of the rotation curve and thickness distribution. The latter is estimated from \citet{EGIS}, where individual vertical cuts along the minor axis were analyzed based on SDSS images in the $r$ band. 

We develop the model with a set of constraints 
similar to the Model 6 described above. Two additional constraints are the use of the halo-to-disk ratio and the gas disk central surface brightness as free parameters.
The former parameter is introduced because we don't use equation (\ref{MhMdhz}) anymore. Note that the latter parameter is highly degenerated with the stellar surface density because of the similar assumed functional form for the gas radial distribution. In the frames of the assumptions the "disk surface density" can be referred to the stars and gas summed together.  
As a result, we obtained qualitatively similar parameters in our modeling here and in the Model 6: this galaxy indicates a rather small halo-to-disk scale ratio and high halo-to-disk mass. Qualitatively, these results
also agree with the pseudo-isothermal halo case in the 
modeling by \citet{banerjee17}. 

\section{Discussion}

Our observations and modeling demonstrate a noticeable dependence of 
STGs properties of their color. The color $(r-i)_0$ correlates with the  
rotation curve maximum (Fig.~\ref{fig3}) and with estimated vertical velocity dispersion of
stars (Fig.~\ref{fig8}). The dark halo scale normalized by the disk scale length 
is shorter in the red STGs (Fig.~\ref{fig8}), although a majority of the blue and almost all red 
STGs have the $a_{\rm halo}/h_d$ ratio less than 2, in agreement with \citet{banerjee13,banerjee17}. 
Most of considered galaxies are
dark matter dominated, and their dark halo mass also correlates with galactic
color (Fig.~\ref{fig11}), as it would be expected from Fig.~\ref{fig3}.

Results of our modeling confirm the conclusion by B17 of a non-uniformity of 
superthin galaxies. The blue STGs with $(r-i)_0 \, \le$ 0.24 mag look dynamically under-evolved: 
Fig.~\ref{fig8} shows that their stellar vertical velocity dispersion is close to 
that expected in the star forming galaxies (see e.g. \citet{kennicutt89,mogotsi16}), which
has been noticed for two superthin galaxies \citep{vdKruit01}. 
At the same time, the central surface density of 
disks in these galaxies is very low, under 120~\Msunpc{} (Fig.~\ref{fig7}). 
The blue STGs preserved themselves from different internal and external factors that 
increase the vertical stellar velocity dispersion: mergers 
\citep{walker96,velazquez99,qu11}
and growth of internal non-axisymmetric instabilities 
\citep{barbanis67,kh10,saha14,grand16}.
As noticed by B17, the blue STGs stay away from busy cosmological environment, like filaments,
which is also typical for LSB galaxies \citep{mo94,rosenbaum04,kautsch09,rosenbaum09}.

The red STGs indicate a  
higher stellar velocity dispersion (but still less than in Milky Way-like galaxies). 
Their disks look thin mostly due to their large disk scale length.
In this sense they resemble the giant LSB galaxies, whose disks may form by continuous 
accretion events rather than a one-epoch assembly \citep{hagen16}. 

In both cases a massive (in the sense of the high 
dark-to-light mass ratio) dark halo
helps suppress the internal instabilities and keeps the disks thin \citep{zasov91,bm02,kautsch09,kh10,ghosh14}.
A central mass concentration like a massive compact bulge or compact halo
helps stabilize the stellar disks against non-asymmetric instabilities and prevents the disks from increasing their thickness \citep{sotnikova05,sotnikova06,kurapati18}.

The role of rotation in shaping the STGs should be essential:
\citet{jadhav19} noticed that the spin angular momentum parameter 
$\lambda$ \citep{peebles69,dalcanton97,bullock01} for 
stars is significantly higher in the STG  
with respect to the normal galaxies, similar to LSBs \citep{maccio07,perezmontano19}.
In addition to a relatively high mass, a parameter of the internal structure of dark 
halo, such as the
halo scale length, is assumed to be a key factor that shapes and preserves the STGs. Thus,
\citet{banerjee13,kurapati18} found that the halo scale is short (under two stellar disk 
scale length) in a few STGs, whereas the regular high surface brightness galaxies have
longer dark halo scales \citep{gentile04,narayan05,banerjee08,kh10}.
\citet{dipaolo19} found that the halo-to-disk scale ratio $R_{halo} / R_{disk}$ anti-correlates with the dark halo
compactness. Our results are in agreement with the short dark halo scale in most
of STGs: Fig.~\ref{fig9} shows that more than a half of galaxies have 
$a_{halo}/h_d$ under 2. At the same time, the blue STGs in our sample show higher
$a_{halo}/h_d$ values (Fig.~\ref{fig9} and \ref{fig10}), which  suggests that they have
more compact dark halos, according to \citet{dipaolo19}. 
This also is in agreement with simulations by \citet{jiang19}.
Note also that the least scale ratios $a_{halo}/h_d$ correspond to the red STGs (Fig.~\ref{fig9}).
They also have massive halos and highest dark-to-luminous mass ratios (Fig.~\ref{fig11}).

We employ the definitions of the spin angular momentum  $\lambda$ from \citet{bullock01,jiang19} to
estimate the  parameter $\lambda_{\rm gal}$ for the disks in our galaxies. 
The mean $\lambda_{\rm gal}$ for the disk is 0.078,
which is some three times higher than in typical galaxies \citep{jiang19}. 
The blue and red STG groups show similar mean $\lambda_{\rm gal}$.

\citet{dalcanton04} noticed that small galaxies with the rotational velocity under 120~\kms{} show a 
lack of the visible dust layer. Our Figures \ref{fig3} and \ref{fig8} show that small STGs
bluer than $(r-i)_0 = 0.24$ mag have their rotational velocities under 120--125~\kms{}
and the vertical stellar velocity dispersion close to that in the gas. It means their disks of 
stars and dust have similar thickness, therefore the dust layer cannot be clearly seen, 
which naturally explains the transition from galaxies with organized dust lanes to those without them at this specific value of the rotational velocities. 

\subsection{Limitations of the Models and the Further Modeling}
Detailed modeling of galaxies is a complex and time consuming task. We decided for an efficient, powerful, and established approach. Our modeling is data driven by adapting a specific vertical distribution of stellar density following 
\citet{vdKruit81,bottema93}. This enables the active comparisons of derived structural parameters in large catalogs, such as \citet{EGIS}. 
The exact solution of Poisson and hydrostatic vertical equilibrium equations 
\citep{bahcall84} generates corrections to the disk vertical density profile in eq. (\ref{rhodisk}) for halo gravitation. As shown by \citet{bahcall84}, this correction is small.

Solving the Poisson and hydrostatic equilibrium equations as in \citet{narayan02,banerjee13,aditya20} and comparing the resulting disk thickness radial distribution with observations has several more degrees of freedom to explore: from testing different density distributions in the dark halo to introducing the density-luminosity constraints. We postpone a more complicated analysis for a future publication. In this work we describe our data-driven procedure in eqs. (5-11) that employs only one scaling relation described by eq. (\ref{MhMdhz}). It is conceived from results of N-body simulations, and proved on both HSB and LSB galaxies therein. Note that two our models developed above do not use eq. (\ref{MhMdhz}) at all. 
Our results show that our controlled modeling facilitates the use in our large set of galaxies with existing catalog data. To further investigate the effectiveness of our model, we plan to compare the different approaches in a specialized upcoming article. 

\subsection{Comparison With Published Results}

Our sample has three STGs studied in the past: UGC~7321, IC~2233, and UGC~9242. 
Besides the first object investigated in detail \citep{matthews00,uson03,banerjee17}, 
IC~2233 was modeled by \citet{banerjee17}. 
Spectra of UGC~9242 were published by \citet{goadroberts81}.
One more galaxy observed by us, UGC~711, was mentioned by \citet{mendelowitz00}, but the 
publication does not have the actual data available for comparison. 



IC~2233:  \citet{banerjee17} employs the \citet{salo15} two-disk fitting to the Spitzer NIR photometry for this galaxy.
Our optical and NIR disk scales are close to the published data. Our ionized gas rotation curve 
maximum is close to that obtained by  \citet{matthews08}. Our rotation curve is limited 
by the 
spectrograph's long slit size, and its spacial extension is 30 per cent smaller than the published HI curve.
This galaxy indicates a non-standard rotation curve affected by small-scale instabilities  \citep{matthews08b}.
Our modeling is not conclusive, and the methods of modeling
applied by us yield either extremely
long or extremely short halo scales.  The rotation curve shape is not smooth for this galaxy, and 
a custom-made model with trials of different mass profiles for the bulge and disk for this galaxy
is required to reproduce its rotation curve better. 

UGC~7321. A state-of-art modeling of the prototype superthin galaxy UGC~7321
was performed by \citet{matthews00,uson03,matthews01,banerjee13}.
We use a shorter distance to the galaxy with respect to \citet{matthews00}. 
Our optical scale length is slightly longer than that from \citet{matthews00} (56 versus 43 arcsec),
while the H band scale length is shorter (21 versus 41 arcsec). The scale heights 
we use are comparable to the published if converted to the exponential density distribution
in the vertical direction. 
The HI rotation curve obtained by \citet{uson03} is twice as more extended than our \Halpha{}
curve, but the latter reaches the distance where the curve becomes flat. 

UGC~9242: the major axis spectrum was published by \citet{goadroberts81}, and it is in agreement
with the spectra we obtained. Our rotation curve is slightly more extended (by 10 per cent) than the published.

\section{Summary}

We obtain spectra of \Nsp superthin galaxies with the resolution of R$\sim$5000
and perform NIR observations of \Nnir STGs from that sample on the 3.5m telescope
at the Apache Point Observatory. 
The combination of our spectroscopy, NIR photometry, published NIR data \citep{nir}, and 
SDSS photometry allows us to develop a set of constrained models taking into account the vertical structure of stellar disks. 
This enables us to study the dark matter halo parameters and 
dynamical status in the largest, uniform sample of superthin galaxies.

We find that the sample of superthin galaxies is not uniform, and can be subdivided
by the blue, $(r-i)_0 \le 0.24$~mag, and red, $(r-i)_0 > 0.24$~mag, STGs.
The blue superthin galaxies are mostly dynamically under-evolved. 
Their stellar vertical velocity dispersion is as low as that in the gas. The red STGs
show higher vertical velocity dispersion, in combination with a very large disk scale length. 

The surface brightness in the disks of the superthin galaxies is constrained by the minimum
vertical stellar velocity dispersion that cannot be lower than that of the galactic
star forming gas. In addition, there may be a limiting central surface density, which is 
of the order of 50~\Msunpc{} from our simplified toy model. 

Most of our STGs are dark matter dominated. Their rotational velocity and dark halo mass correlates 
with galactic color. The blue STGs also have less compact dark halos than  the red STGs, 
whereas the galaxies in both color groups have their halo-to-disk scales ratios under two.
 
\begin{table*}
\centering
\caption{Parameters of the galaxies obtained in
Model 6. The columns show the name, central 
surface density (in \Msunpc), fitted disk scale
length (kpc), halo scale (kpc), dark halo mass, 
disk mass, and the best fitting chi-square value.}
\begin{tabular}{lllllll} 
\hline 
Name & $\Sigma_0$ & hfit & ahalo & log$M_{halo}$ & log$M_d$ & chi**2 \\
\hline
\hline
EON\_105.825\_13.464  &162.&  7.67&  8.03& 11.22& 10.80&  3.215e+00\\
EON\_11.160\_-11.189  &222.&  4.37& 45.19& 10.27& 10.43&  8.352e+00\\
EON\_115.887\_31.535  &147.&  2.28& 49.96&  9.58&  9.89&  1.000e+00\\
EON\_116.146\_18.328  &269.&  4.83&  0.14& 10.65& 10.60&  9.946e+00\\
EON\_118.918\_28.743  &122.&  2.16&  9.05&  0.00&  9.73&  5.063e+00\\
\multicolumn{7}{c}{... Table is published in its entirety}\\
\multicolumn{7}{c}{in the electronic edition. ...}\\
\hline
\end{tabular}
\label{tab6}
\end{table*}



\section*{Acknowledgements}
The project is partly supported by RScF grant 19--12--00145.
VPR and AVA acknowledge the support by RFBR grant 19--32--50129 for an analysis of the photometry of the superthin galaxies.
The database support was provided in framework of RFBR grant 19--32--90244.

We thank anonymous referee whose comments and valuable suggestions improved the paper. 
Based on observations obtained with the Apache Point
Observatory 3.5-meter telescope, which is owned and operated by the
Astrophysical Research Consortium.

We acknowledge the usage of the HyperLeda database\footnote{\url{http://leda.univ-lyon1.fr}} \citep{HyperLeda}.  
This research has made
use of the NASA/IPAC Extragalactic Database (NED) which is operated by the
Jet Propulsion Laboratory, California Institute of Technology, under
contract with the National Aeronautics and Space Administration.

{}


\end{document}